\documentclass[letterpaper, 10 pt, conference]{ieeeconf}  

\IEEEoverridecommandlockouts                              
                                                          
\usepackage{amsmath}
\usepackage{amssymb}
\usepackage{tabularx}
\usepackage{xfrac}
\usepackage[final]{pdfpages}
\usepackage[ruled,vlined,noend,linesnumbered]{algorithm2e}
\usepackage{algpseudocode}
\usepackage{booktabs}
\usepackage[short]{optidef}
\usepackage{soul,xcolor}
\usepackage{siunitx}
\usepackage{import}
\usepackage{bm}
\usepackage{tablefootnote}
\usepackage{physics}

\usepackage{units}
\usepackage{arydshln}
\usepackage{lipsum}

\usepackage{url}
\usepackage[hidelinks]{hyperref}
\usepackage{balance}

\usepackage{graphicx}
\usepackage{comment} 

\usepackage{cite}
\usepackage{mathtools,cuted}

\usepackage[capitalise]{cleveref}

\usepackage{caption}
\usepackage{subcaption}

\usepackage{nomencl}
\makenomenclature

\Crefname{equation}{Eq.}{Eqs.}
\Crefname{figure}{Fig.}{Figs.}
\Crefname{tabular}{Tab.}{Tabs.}
\Crefname{line}{L.}{L.}

\renewcommand{\nompreamble}{The next list describes several symbols that we used throughout the paper and their explanation.}

\newlength\figureheight
\newlength\figurewidth

\newcommand{\Gp}{G_{\text{p}}} 
\newcommand{\Gv}{G_{\text{v}}} 
\newcommand{\Kp}{K_{\text{p}}} 
\newcommand{\Kv}{K_{\text{v}}} 
\newcommand{\Ti}{T_{\text{i}}} 
\newcommand{\Cp}{C_{\text{p}}} 
\newcommand{\Cv}{C_{\text{v}}} 

\newcommand{\vc}[1]{{ \mathrm{#1} }}

\newcommand{\suggestion}{x^*}
\newcommand{\data}{\mathcal{D}}
\newcommand{\domain}{\mathcal{X}}
\newcommand{\tdomain}{\mathcal{T}}
\newcommand{\threshold}{\kappa}
\newcommand{\g}[2]{g_\epsilon^t(#1,#2)}
\newcommand{\task}{\tau}

\newcommand{\xbest}{x^\dagger}
\newcommand{\Kpbest}{K_{\text{p}}^\dagger}
\newcommand{\Kvbest}{K_{\text{v}}^\dagger}
\newcommand{\Tibest}{T_{\text{i}}^\dagger}
\newcommand{\xbestgrid}{x^\dagger_\text{grid}}

\newcommand{\mypar}[1]{\textbf{#1.}}
\newcommand{\safeopt}{\textsc{SafeOPT}}
\newcommand{\stageopt}{\textsc{StageOPT}}
\newcommand{\goose}{\textsc{GoOSE}}

\DeclareMathOperator*{\argmin}{arg\,min}
\newcommand{\Xcal}{{\mathcal{X}}}
\newcommand{\Rbb}{{\mathbb{R}}}

\newboolean{include-notes}
\setboolean{include-notes}{true}

\newcommand{\matteo}[1]{\ifthenelse{\boolean{include-notes}}{{\color{orange} \textbf{Matteo}: #1}}{}}
\newcommand{\christopher}[1]{\ifthenelse{\boolean{include-notes}}{{\color{green} \textbf{Christopher}: #1}}{}}
\newcommand{\alisa}[1]{\ifthenelse{\boolean{include-notes}}{{\color{red} \textbf{Alisa}: #1}}{}}
\newcommand{\andreas}[1]{\ifthenelse{\boolean{include-notes}}{{\color{blue} \textbf{Andreas}: #1}}{}}

\begin{document}


\title{Safe and Efficient Model-free Adaptive Control via Bayesian Optimization}
\author{Christopher K{ö}nig$^{1,*}$, Matteo Turchetta$^{2,*}$, John Lygeros$^{3}$, Alisa Rupenyan$^{1,3}$, Andreas Krause$^{2}$
\thanks{This project has been finded by the Swiss Innovation Agency (Innosuisse), grant Nr. 46716, and by the Swiss National Science Foundation under NCCR Automation.\newline}
\thanks{$^{1}$ Inspire AG, Zurich, Switzerland }\newline

\thanks{$^{2}$ Learning \& Adaptive Systems group, ETH Zurich, Switzerland} 
\thanks{$^{3}$ Automatic Control Laboratory, ETH Zurich, Switzerland }\newline
\thanks{$^*$ The authors contributed equally.}
}
\maketitle


\begin{abstract}
\looseness=-1
 Adaptive control approaches yield high-performance controllers when a precise system model or suitable parametrizations of the controller are available. Existing data-driven approaches for adaptive control mostly augment standard model-based methods with additional information about uncertainties in the dynamics or about disturbances. In this work, we propose a purely data-driven, model-free approach for adaptive control. Tuning low-level controllers based solely on system data raises concerns on the underlying algorithm safety and computational performance. Thus, our approach builds on \goose{}, an algorithm for safe and sample-efficient Bayesian optimization. We introduce several computational and algorithmic modifications in \goose{} that enable its practical use on a rotational motion system. We numerically demonstrate for several types of disturbances that our approach is sample efficient, outperforms constrained Bayesian optimization in terms of safety, and achieves the performance optima computed by grid evaluation. We further demonstrate the proposed adaptive control approach experimentally on a rotational motion system.
\end{abstract}
\section{Introduction}

Adaptive control approaches are a desirable alternative to robust controllers in high-performance applications that deal with disturbances and uncertainties in the plant dynamics. Learning uncertainties in the dynamics and adapting have been explored with classical control mechanisms such as Model Reference Adaptive Control (MRAC) \cite{Chowdhary2015,Grande2014}. Gaussian processes (GP) have been also used to model the output of a nonlinear system in a dual controller \cite{Sbarbaro2005}, while coupling the states and inputs of the system in the covariance function of the GP model. Learning dynamics in an $\mathcal{L}_1$-adaptive control approach has been demonstrated in \cite{gahlawat20a,Fan2019}. 

Instead of modeling or learning the dynamics, the system can be represented by its performance, directly measured from data. Then, the low-level controller parameters can be optimized to fulfill the desired performance criteria. This has been demonstrated for motion systems in \cite{Berkenkamp2, Duivenvoorden, Khosravi2020b,Khosravi2020,Koenig2020}. Such model-free approaches, however, have not been brought to continuous adaptive control, largely because of difficulties in continuously maintaining  stability and safety in the presence of disturbances and system uncertainties, and because of the associated computational complexity. Recently, a sample-efficient extension for safe exploration in Bayesian optimization has been proposed \cite{turchetta2019safe}. In this paper, we further optimize this algorithm to develop a model-free adaptive control method for motion systems. 

\mypar{Contribution} 
In this work, we make the following contributions: (1) we extend the \goose{} algorithm for policy search to adaptive control problems; that is, to problems where constant tuning is required due to changes in environmental conditions. (2) We reduce \goose{}'s complexity so that it can be effectively used for policy optimization beyond simulations. (3) We show  the effectiveness of our approach in extensive evaluations on a real and simulated rotational axis drive, a crucial component in many industrial machines.

A table of symbols can be found in \cref{sec:appendix}.
\section{Related work}
\mypar{Bayesian optimization}
Bayesian optimization (BO) \cite{Mockus} denotes a class of sample-efficient, black-box optimization algorithms that have been used to address a wide range of problems, see \cite{shahriari2015taking} for a review. In particular, BO has been successful in learning high-performance controllers for a variety of systems. For instance, \cite{calandra2016bayesian} learns the parameters of a discrete event controller
for a bipedal robot with BO, while \cite{marco2017virtual}, trades off real-world and simulated control experiments via BO.
In \cite{antonova2020bayesian}, variational autoencoders are combined with BO to learn to control  an hexapod, while \cite{turchetta2019robust} uses multi-objective BO to learn robust controllers for a pendulum.

\mypar{Safety-aware BO} 
Optimization under unknown constraints naturally models the problem of learning in safety-critical conditions, where \textit{a priori} unknown safety constraints must not be violated. 
In \cite{Gardner} and \cite{hernndezlobato2015predictive} safety-aware variants of standard BO algorithms are presented. In \cite{picheny2016bayesian,gramacy2016modeling,ariafar2019admmbo}, BO is used as a subroutine to solve the unconstrained optimization of the augmented Lagrangian of the original problem. While these methods return feasible solutions, they may perform unsafe evaluations. In contrast, the \safeopt{}  algorithm \cite{Sui} guarantees safety at all times. It has been used to safely tune a quadrotor controller for position tracking \cite{Berkenkamp,Berkenkamp2}. In \cite{Duivenvoorden}, it has been integrated with particle swarm optimization (PSO) to learn high-dimensional controllers. Unfortunately, \safeopt{} may not be sample-efficient due to its exploration strategy. To address this, many solutions have been proposed. 
For example, \cite{fiducioso2019safe} does not actively expand the safe set, which may compromise the optimality of the algorithm but works well for the application considered. Alternatively, \stageopt{} \cite{sui2018stagewise} first expands the safe set and, subsequently, optimizes over it. Unfortunately, it cannot provide good solutions if stopped prematurely. \goose{} \cite{turchetta2019safe} addresses this problem by using a separate optimization oracle and expanding the safe set in a goal-oriented fashion only when necessary to evaluate the inputs suggested by the oracle. 
\section{System and problem statement}\label{sec:problem_statement}
In this section, we present the system of interest, its control scheme, and the mathematical model we use for our numerical evaluations. Finally, we introduce the safety-critical adaptive control problem we aim to solve.

\subsection{System and controller}
\label{section:system}
\begin{figure}[h]
\centering
\raisebox{-.5\height}{\includegraphics[width=0.21\textwidth]{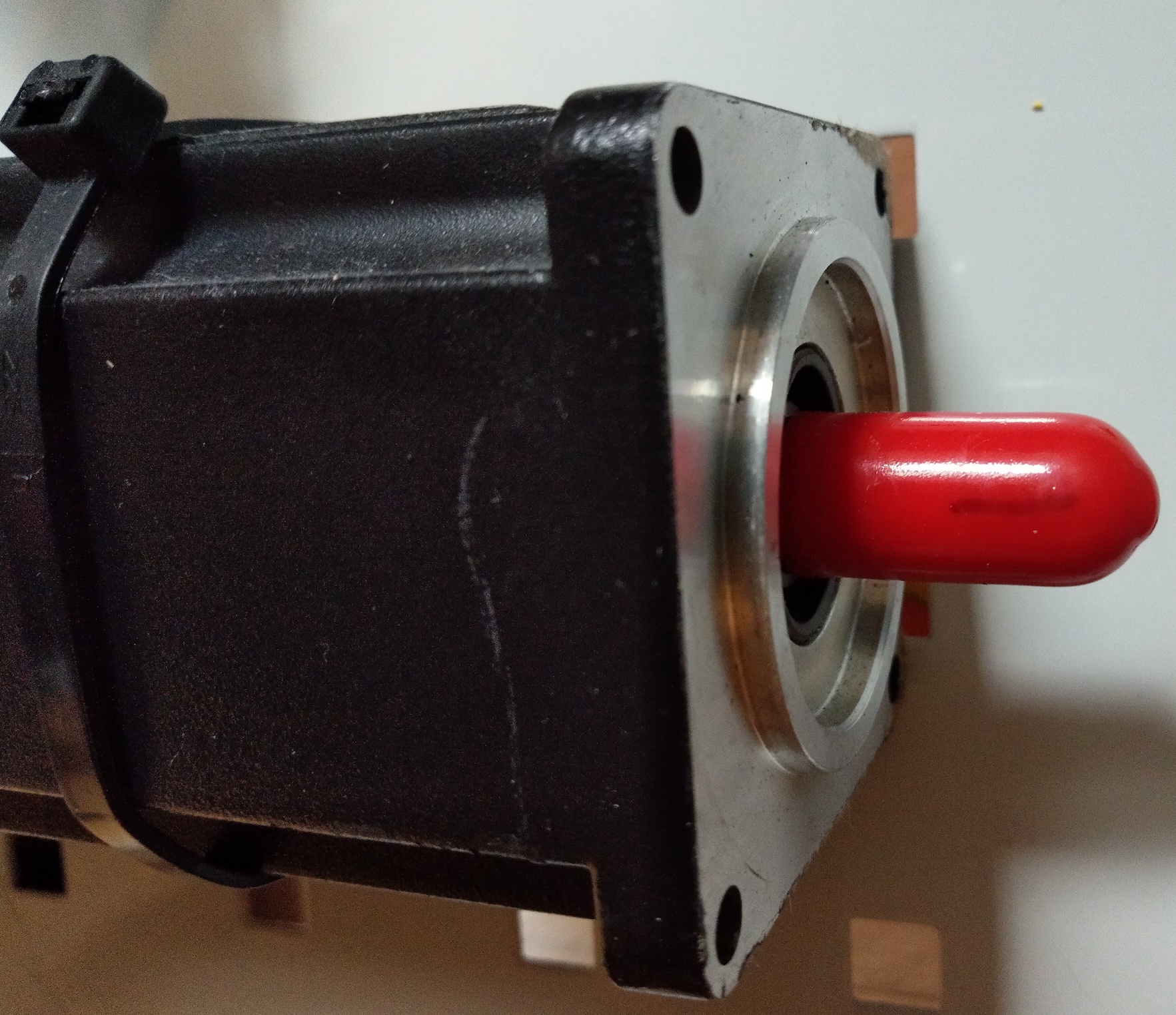}}
\centering
\small\addtolength{\tabcolsep}{-0pt}
\begin{tabular}{ @{}l l l @{} }
\toprule
 Param. & Value & Unit\\
 \midrule
$m$  & $0.0191$ & $\mathrm{kgm^2}$\\
$b$   & $30.08$ & $\mathrm{kgm^2/s}$\\
$c_1$ & $1.78e-3$ & $\mathrm{Nm}$\\
$c_2$ & $0.0295$ & $\mathrm{Nm/rad}$\\
$c_3$ & $0.372$ & $\mathrm{rad}$\\
$c_4$ & $8.99e-3$ & $\mathrm{Nm}$\\
$c_5$ & $0.11$ & $\mathrm{rad}$\\
\bottomrule
\end{tabular}
\caption{Rotational axis drive (left) and its parameters (right).}
\label{fig:rotational_drive_with_parameters}
\end{figure}

The system of interest is a rotational axis drive, a positioning mechanism driven by a  synchronous 3 phase permanent magnet AC motor equipped with encoders for position and speed tracking (see \cref{fig:rotational_drive_with_parameters}). 
Such systems are routinely used as components in the semiconductor industry, in biomedical engineering, and in photonics and solar technologies. 

We model the system as a combination of linear and nonlinear blocks, where the linear block is modeled as a damped single mass system, following \cite{Khosravi2020}:
\begin{equation}\label{eqn:Gs}
G(s) := \left[\Gp(s), \Gv(s) \right]^\top=
\left[\frac{1}{ms^2 + bs},\frac{1}{ms + b}\right]^\top,    
\end{equation}
where $\Gp(s)$ and $\Gv(s)$ are the transfer functions respectively from torque to angular position and torque to angular velocity, $m$ is  the moment of inertia, and  $b$ is the rotational damping coefficient due to the friction. The values of $m$ and $b$ are obtained via least squares fitting and shown in  \cref{fig:rotational_drive_with_parameters}. 

\begin{figure}[h]
\includegraphics[width=0.5\textwidth]{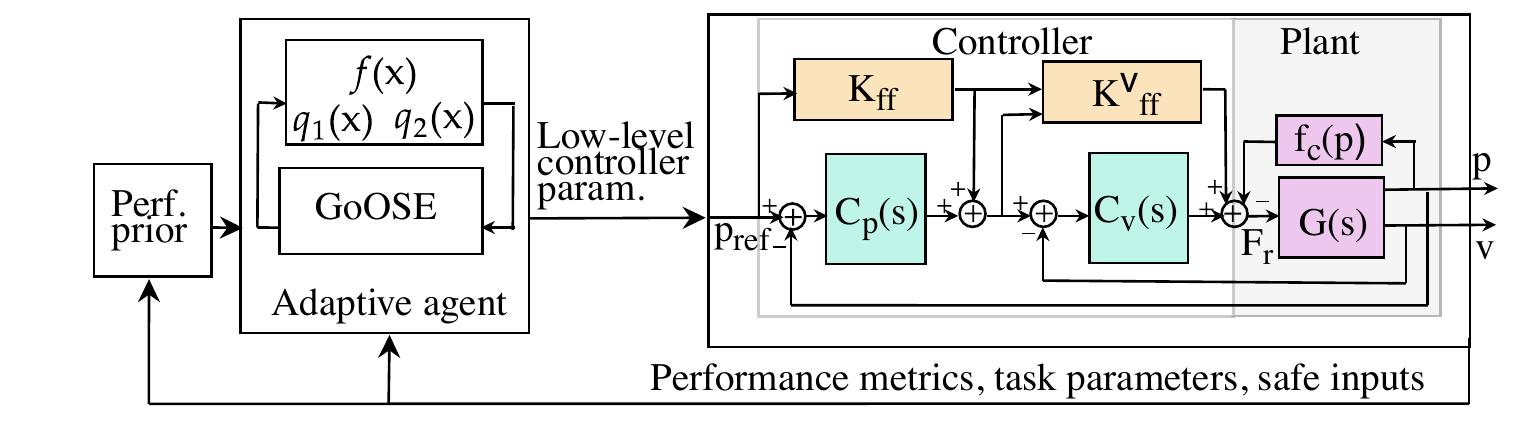}
\caption{Scheme of the proposed BO-based adaptive control.}
\vspace{-0.2cm}
\label{fig:Scheme}
\end{figure}

Next, we introduce the model for the nonlinear part of the dynamics $f_\vc{c}$, which we subtract from the total torque signal, see \cref{fig:Scheme}.
The nonlinear cogging effects due to interactions between the permanent magnets of the rotor and the stator slots
\cite{Villegas} are modelled using a Fourier truncated expansion as $f_\vc{c}(p) = c_1 + c_2p +  
\sum_{k=1}^{n} c_{2 k+2}\!\ \sin\!\big{(}\frac{2 k \pi}{c_{3}}p+c_{2 k+3}\big{)}\, $
where $p$ is the position, $c_1$ is the average thrust torque, $c_2$ is the gradient of the curve, $c_3$ is the largest dominant period described by the angular distance of a pair of magnets, $n$ is the number of modelled frequencies, and, $c_{2k+2}$ and $c_{2k+3}$ are respectively the amplitudes and the phase shifts of the sinusoidal functions, for $k=1, 2,\ldots,n$.
The parameters 
$\vc{c}:=[c_1,\ldots,c_{2n+3}]$
are estimated using least squares error minimization between the modelled cogging torque signals and the measured torque signal at constant velocity to cancel the effects of linear dynamics. The estimates of the parameters are shown in \cref{fig:rotational_drive_with_parameters}. To model the noise of the system, zero mean white noise with 6.09e-3 Nm variance is added to the torque input signal of the plant.

The system is controlled by a three-level cascade controller shown in Figure \ref{fig:Scheme}. 
The outermost loop controls the position with a P-controller $\Cp(s)=\Kp$, and the middle loop controls the velocity with a PI-controller $\Cv(s) = \Kv(1+\frac{1}{\Ti s})$. 
The innermost loop controls the current of the rotational drive. It is well-tuned and treated as a part of the plant $G(s)$. Feedforward structures are used to accelerate the response of the system. Their gains are well-tuned and not modified during the retuning procedure. However, in our experiments, we perturb them to demonstrate that our method can adapt to new operating conditions by adjusting the tunable parameters of the controller.

\subsection{Adaptive control approach} \label{section:adaptive_control_problem}
In this section, we present the adaptive control problem we aim to solve. In particular, our goal is to tune the parameters of the cascade controller introduced in \cref{section:system} to maximize the tracking accuracy of the system, following \cite{Khosravi2020}. Let $\domain$ denote the space of admissible controller parameters, $x=\left[\Kp, \Kv, \Ti\right]$, and let $f: \domain\to\Rbb$ be the objective measuring the corresponding tracking accuracy. In particular, we define $f$ as the position tracking error averaged over the trajectory induced by the controller $f(x) = \frac{1}{N}\sum_{i = 1}^{N} \abs{p_i^{\mathrm{err}}(x)}$, where $p_i^{\mathrm{err}}$ is the deviation from the reference position at sampling time $i$. Crucially, $f$ does not admit a closed-form expression even when the system dynamics are known. However, for a given controller $x \in \Xcal$, the corresponding tracking accuracy $f(x)$ can be obtained experimentally. Notice that $f$ can be extended to include many performance metrics, to  minimise oscillations, or to reduce settling time, as in \cite{Khosravi2020,Khosravi2020b}.

\looseness=-1
In practice, we cannot experiment with arbitrary controllers while optimizing $f$ due to safety and performance concerns. Thus, we introduce two constraints that must be satisfied at all times. The first one is a safety constraint $q_1(x)$  defined as the maximum of the fast Fourier transform (FFT) of the torque measurement in a fixed frequency window. The second one is a tracking performance constraint   $q_2(x)=|p^{\mathrm{err}}(x)|_{\infty}$ defined as the infinity norm of the position tracking error. Finally, in reality, the system may be subject to sudden or slow disturbances such as a change of load or a drift in the dynamics due to components wear. Therefore, our goal is to automatically tune the controller parameters of our system to maximize tracking accuracy for varying operating conditions that we cannot control, while never violating safety and quality constraints along the way.
\section{Background}\label{sec:backgound}
\mypar{Gaussian processes}
A Gaussian process (GP) \cite{Rasmussen} is a distribution over the space of functions commonly used in non-parametric Bayesian regression. It is fully described by a mean function $\mu:\domain \rightarrow \mathbb{R}$, which, w.l.o.g, we set to zero for all inputs $\mu(x)=0,~\forall x\in \domain$, and a kernel function $k:\domain \times \domain \rightarrow \mathbb{R}$. Given the data set $\data=\{(x_i,y_i\}_{i=1}^t$, where $y_i=f(x_i)+\varepsilon_i$ and $\varepsilon_i\sim \mathcal{N}(0,\sigma^2)$ is zero-mean i.i.d. Gaussian noise, the posterior belief over the function $f$ has the following mean, variance and covariance:
\begin{align}
    &\mu_t(x) = \mathbf{k}_t^\top(x) (\mathbf{K}_t + \sigma^2 \mathbf{I})^{-1} \mathbf{y}_t, \label{eq:GPmean}\\
    &k_t(x,x')=k(x,x') - \mathbf{k}_t^\top(x) (\mathbf{K}_t + \sigma^2 \mathbf{I})^{-1} \mathbf{k}_t(x'), \label{eq:GPcovar}\\
    &\sigma_t(x)=k_t(x,x)\label{eq:GPvar},
\end{align}
where ${\mathbf{k}_t(x)=(k(x_1,x), \dots, k(x_t,x))}$, $\mathbf{K}_t$ is the positive definite kernel matrix $[k(x,x')]_{x,x' \in \data_t}$, and $\mathbf{I} \in \mathbb{R}^{t \times t}$ denotes the identity matrix. In the following, the superscripts $f$ and $q$ denote GPs on the objective and on the constraints.

\mypar{Multi-task BO}
In multi-task BO, the objective depends on the  extended input $(x,\task) \in \domain \times \tdomain$, where $x$ is the variable we optimize over and $\task$ is a task parameter set by the environment that influences the objective. To cope with this new dimension, multi-task BO adopts kernels of the form $k_\text{multi}((x,\task),(x',\task')) = k_{\task}(\task,\task')\otimes k(x,x')$, where $\otimes$ denotes the Kronecker product. This kernel decouples the correlations in objective values along the input dimensions, captured by $k$, from those across tasks, captured by $k_\task$ \cite{Swersky2013}.

\mypar{\goose{}}
\looseness=-1
\goose{} extends any standard BO algorithm to provide high-probability safety guarantees in presence of \textit{a priori} unknown safety constraints. It builds a Bayesian model of the constraints from noisy evaluations based on GP regression. It uses this model to build estimates of two sets: the \emph{pessimistic} safe set, which contains inputs that are safe, i.e., satisfy the constraints, with high probability and the \emph{optimistic} safe set that contains inputs that could potentially be safe. At each round, \goose{} communicates the optimistic safe set to the BO algorithm, which returns the input it would evaluate within this set, denoted as $\suggestion$. If $\suggestion$ is also in the pessimistic safe set, \goose{} evaluates the corresponding objective. Otherwise, it evaluates the constraints at a sequence of provably safe inputs, whose choice is based on a heuristic priority function, that allow us to conclude that $\suggestion$ either satisfies or violates the constraints with high probability. In the first case, the corresponding objective value is observed. In the second case, $\suggestion$ is removed from the optimistic safe set and the BO algorithm is queried for a new suggestion. Compared to \cite{sui2018stagewise,Sui}  \goose{} achieves a higher sample efficiency \cite{turchetta2019safe} while compared to \cite{Gardner,hernndezlobato2015predictive,picheny2016bayesian,gramacy2016modeling,ariafar2019admmbo}, it guarantees safety at all times with high probability, under  regularity assumptions.

\mypar{\goose{} assumptions}
\looseness=-1
To infer  constraint and objective values of inputs before evaluating them, \goose{} assumes that these functions belong to a class of well-behaved functions, i.e., functions with a bounded norm in some reproducing kernel Hilbert space (RKHS) \cite{scholkopf2018learning}. Based on this assumption, we can build well-calibrated confidence intervals over them. Here, we present these intervals for the safety constraint, $q$ (the construction for $f$ is analogous).
Let $\mu^q_t(x)$ and $\sigma^q_t(x)$ denote the posterior mean and standard deviation of our belief over $q(x)$ computed according to \cref{eq:GPmean,eq:GPcovar}. We recursively define these monotonically increasing/decreasing lower/upper bounds for $q(x)$: $l^q_t(x)=\max (l^q_{t-1}(x), \mu^q_{t-1}(x) - \beta^q_{t-1}\sigma^q_{t-1}(x))$ and $u^q_t(x)=\min (u^q_{t-1}(x), \mu^q_{t-1}(x) + \beta^q_{t-1}\sigma^q_{t-1}(x))$. The authors of \cite{chowdhury2017kernelized,srinivas2009gaussian} show that, for functions with bounded RKHS norm, an appropriate choice of $\beta^q_t$ implies that $l_t^q(x)\leq q(x) \leq u^q_t(x)$ for all $t \in \mathbb{R}^+$ and $x\in \domain$. However, in practice, a constant value of $3$ is sufficient to achieve safety \cite{Berkenkamp,Berkenkamp2,turchetta2019safe}.

To start collecting data safely, \goose{} requires knowledge of a set of inputs that are known \textit{a priori} to be safe, denoted as $S_0$. In our problem, it is easy to identify such set by designing conservative controllers for simplistic first principle models of the system under control.
\section{\goose{} for adaptive control}\label{sec:method} 
\looseness=-1
In adaptive control, quickly finding a safe, locally optimal controller in response to modified external conditions is crucial.
In its original formulation, \goose{} is not suitable to this problem for several reasons: (\textit{i}) it assumes knowledge of the Lipschitz constant of the constraint, which is unknown in practice, (\textit{ii}) it relies on a fine discretization of the domain, which is prohibitive in large domains, (\textit{iii}) it explicitly computes the optimistic safe set, which is expensive, (\textit{iv}) it does not account for external changes it cannot control. In this section, we address these problems step by step and we present the resulting algorithm.

\begin{algorithm} 
\SetAlgoLined
    \textbf{Input}: Safe seed $S_0$, $f\sim \mathcal{GP}(\mu^f,k^f;\theta^f)$, $q\sim \mathcal{GP}(\mu^q,k^q;\theta^q)$, $\task = \task_0$\;
    Grid resolution: $\Delta x$ s.t. $k^q(x,x+\Delta x)(\sigma^q_\eta)^{-2}$ = 0.95 \label{alg:grid_res}
    \SetInd{0.2em}{0.2em}
    \While {machine is running}{
    $S_t\gets \{x \in \domain:u^q_t(x,\task) \leq\ \threshold\}$\; \label{alg:Spess}
    $L_t \gets \{x \in S_t: \exists z \notin S_t$, with $d(x,z) \leq \Delta x\}$\; \label{alg:L}
    $W_t \gets \{x \in L_t: u^q_t(x,\task) - l^q_t(x,\task)\geq \epsilon\}$\; \label{alg:W}
    $\suggestion \gets \texttt{PSO}(S_t,W_t)$\; \label{alg:get_suggestion}
    \SetInd{0.2em}{0.2em}
    \If {$|f(\xbest(\task)) - l^f_t(\suggestion,\task)| \geq \epsilon_\text{tol}$ \label{alg:stopping}}{
    \lIf{$u^q_t(\suggestion,\task)  \leq\ \threshold$ \label{alg:safe_evaluation}}{
        evaluate $f(x^*),q(x^*)$, 
        update $\task$}\label{alg:update_opt} 
        \SetInd{0.2em}{0.2em}
        \Else{ \SetInd{0.2em}{0.3em} 
        \While{$\exists x\in W_t,~\textrm{s.t.}~\g{x}{\suggestion} > 0$}
        {
        $x_w^* \gets \argmin _{x \in W_t} d(\suggestion,x)$ s.t. $\g{x}{\suggestion} \neq 0$\; \label{alg:get_expander}
        Evaluate $f(x_w^*),q(x_w^*)$, 
        update $\task$, $S_t$, $L_t$, $W_t$\; \label{alg:evaluate_expander} \label{alg:update_exp}
    }
   }
  }
  \Else{set system to $\xbest(\task)$, update $\task$}
 }
 \caption{\goose{} for adaptive control} \label{alg:goose}
\end{algorithm}
\begin{algorithm}
\SetAlgoLined
 \textbf{Input}: Safe set $S_t$, Boundary $W_t$, $\#$ particles $m$\;
 $p_i\sim \mathcal{U}(S_t),~v_i\sim \pm\Delta x,~i=1,\cdots,m$\; \label{alg:pso_init} 
\SetInd{0.2em}{0.2em}
 \For{$j\gets 0$ \KwTo $J$}{ \SetInd{0.2em}{0.3em}
    \For{$i\gets 0$ \KwTo $m$}
    {
    $cond \gets u^q_t(p_i^{j},\task) \leq \threshold \lor 
    \exists x \in W_t: \g{x}{p_i^j}>0$\;
    \label{alg:pso_optimistic_condition}
    $cond \gets cond \land l^f_{t}(p_i^{j},\task) < z_i^{j-1}$\; \label{alg:pso_imptovement}
    \algorithmicif~ $cond$ \algorithmicthen~ $z_i^j \gets l^f_{t}(p_i^{j},\task)$ \algorithmicelse~ $z_i^j \gets z_i^{j-1}$\;}
    \label{alg:pso_z_update}
    $\overline{z}^j\gets \argmin_{i=1,\ldots,m} z_i^j$\; \label{alg:pso_overz_update}
    $r_{1,2} \sim \mathcal{U}([0,2])$\;
    $v_i^{j+1} \gets \alpha^j v_i^j + r_1 (z_i^j - p_i^j) + r_2 (\overline{z}^j - p_i^j),~\forall i$\; \label{alg:pso_v_update}
    $\tilde{p}_i^{j+1}\gets p_i^{j} + v^j_i,~\forall i$\; \label{alg:pso_ptilde_update} 
    }
 \textbf{return} $\overline{z}^J$
 \caption{PSO}\label{alg:pso}
\end{algorithm}

\begin{figure*} [t]
\begin{subfigure}[c]{0.23\textwidth}
\includegraphics[width=1\textwidth]{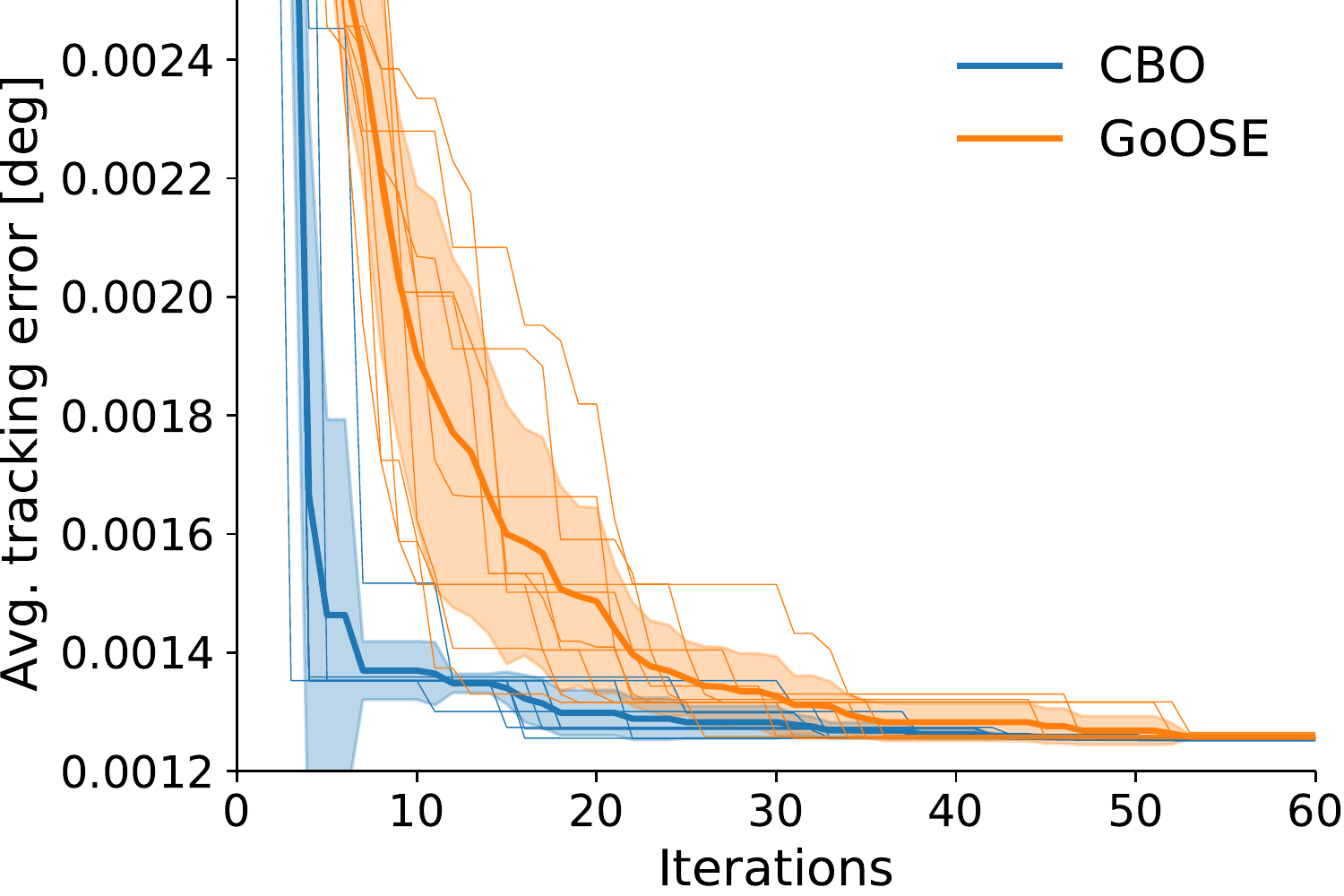}
\end{subfigure}
\begin{subfigure}[c]{0.23\textwidth}%
\includegraphics[width=1\textwidth]{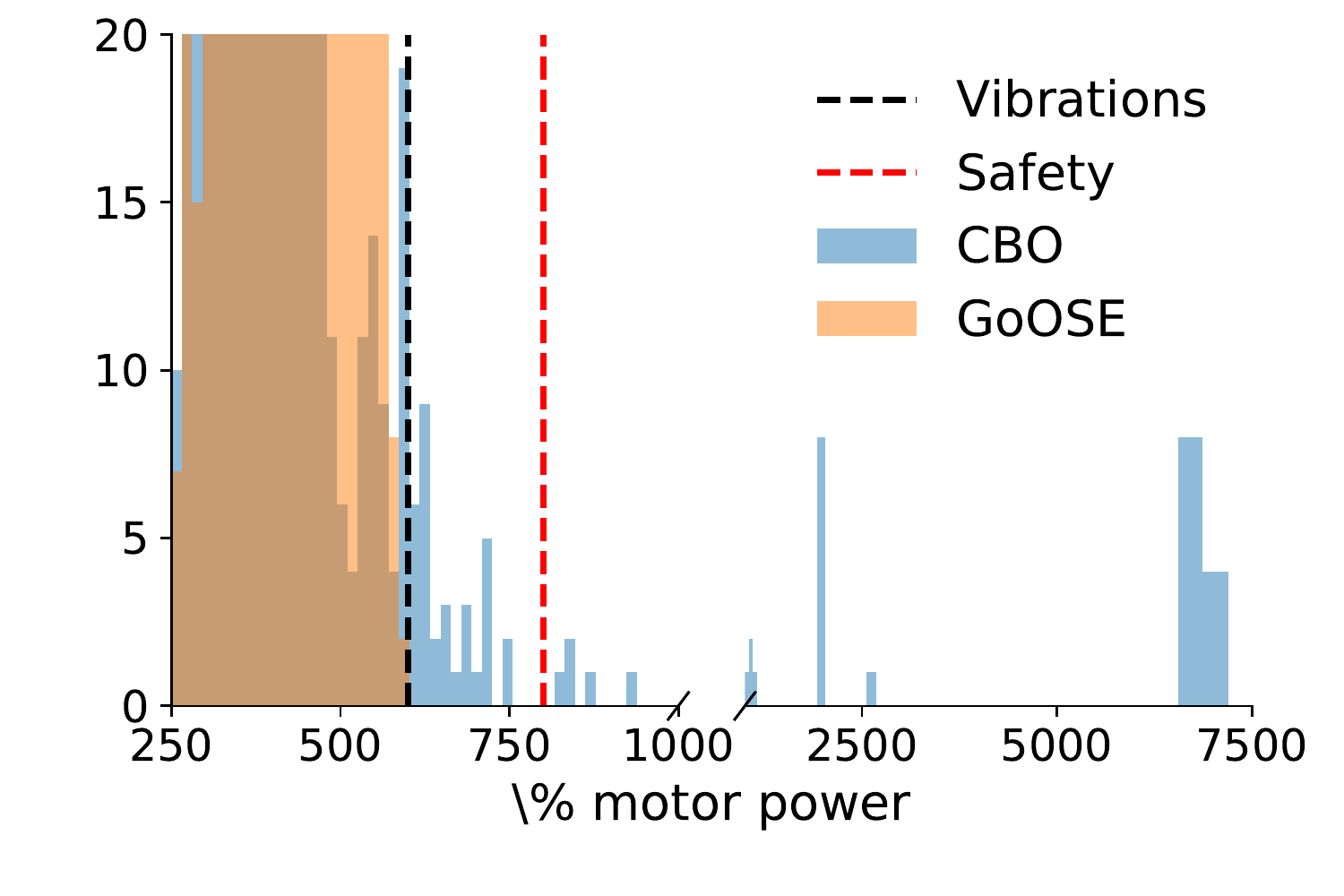} %
\end{subfigure}
\begin{subfigure}[c]{0.23\textwidth}
\includegraphics[width=1\textwidth]{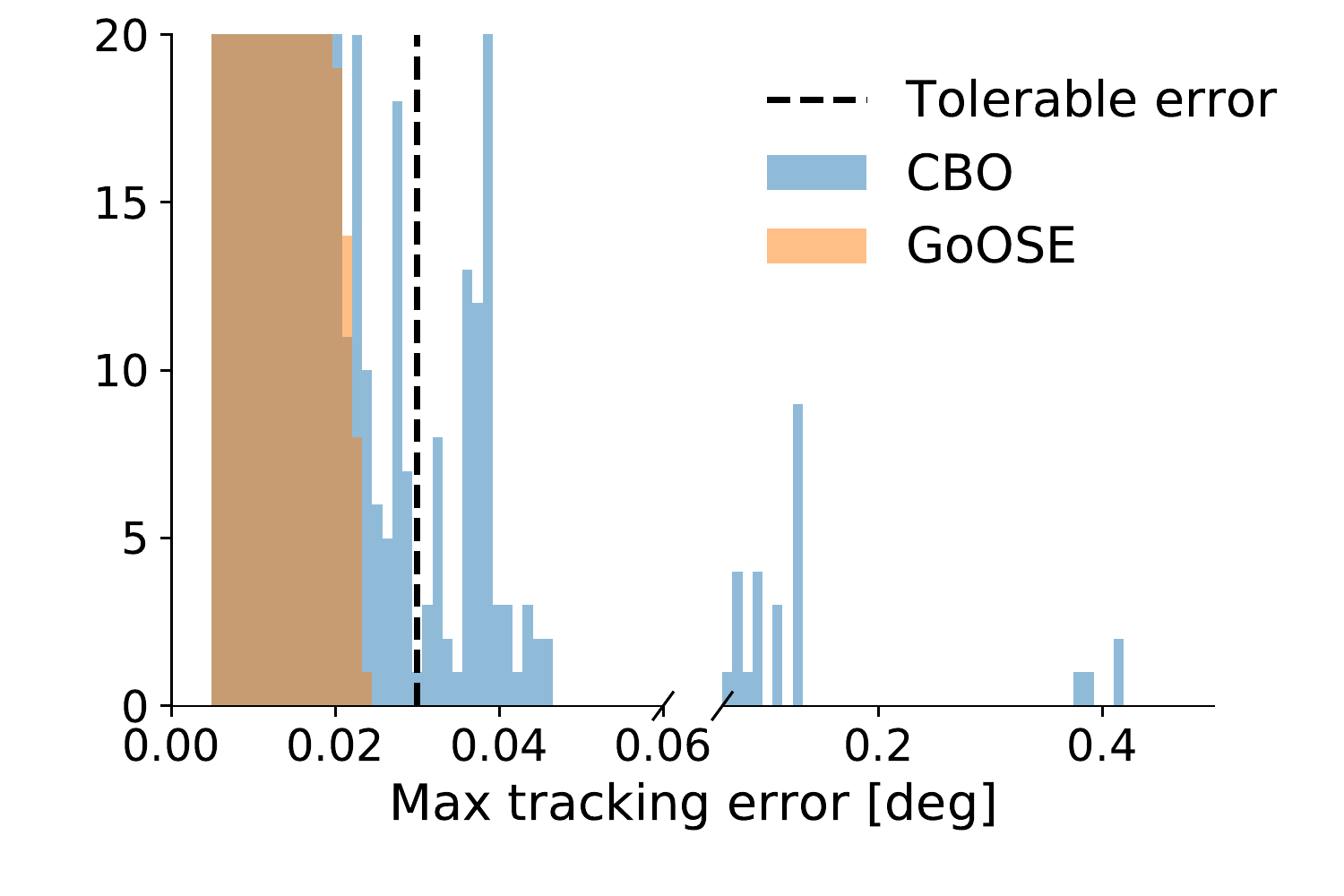}
\end{subfigure}
 \addtolength{\tabcolsep}{-6pt}
 \footnotesize
    \begin{tabular}{ @{}l c c c@{} }\toprule
    & GoOSE & CBO\\\midrule
    It. to conv. & $35.5$ & $18$\\
    $f(\xbestgrid)$  & $1.25$
    & $1.25$\\
    $(\Kpbest, \Kvbest, \Tibest)$ & $(50,0.10,1)$ & $(50,0.09,1)$\\
    \# Violations & $0$ & $5$ \\
    Max. violation & (-,-) & $(6587,0.128)$ \\
   \bottomrule
    \end{tabular}
\caption{\small Comparison of \goose{} and CBO for 10 runs of the stationary control problem. On the left, we see the cost for each run and its mean and standard deviation. The center and right figures show the constraint values sampled by each method.  \goose{} reaches the same solution as CBO (table), albeit more slowly (left). However, CBO heavily violates the constraints (center-right).}
\vspace{-0.2cm}
\label{fig:cbo_comparison}
\end{figure*}
\looseness=-1
\mypar{Task parameter}
In general, the dynamics of a controlled system may vary due to external changes. For example, our rotational axis drive may be subject to different loads or the system components may wear due to extended use. As dynamics change, so do the optimal controllers. In this case, we must adapt to new regimes imposed by the environment. To this end, we extend \goose{} to the multi-task setting presented in \cref{sec:backgound} by introducing a task parameter $\task$ that captures the exogenous conditions that influence the system's dynamics, and by using the kernel introduced in \cref{sec:backgound}. To guarantee safety, we assume that the initial safe seed $S_0$ contains at least one safe controller for each possible task. 

\looseness=-1
\mypar{Lipschitz constant}
\goose{} uses the Lipschitz constant of the safety constraint $L_q$, which is not known in practice, to compute the pessimistic safe set and an optimistic upper bound on constraint values. For the pessimistic safe set, we adopt the solution suggested in \cite{Berkenkamp2} and use the upper bound of the confidence interval to compute it (see \cref{alg:Spess} of \cref{alg:goose}). While pessimism is crucial for safety, optimism is necessary for exploration. To this end, \goose{} computes an optimistic upper bound for the constraint value at input $z$ based on the lower bound of the confidence interval at another input $x$ as $l^q_t(x,\task)+L_q d(x,z)$, where $d(\cdot,\cdot)$ is the metric that defines the Lipschitz continuity of $q$.
Here, we approximate this bound with $l^q_t(x,\task)+\|\mu^{\nabla}_t(x,\task)\|_{\infty}d(x,z)$, where $\mu^{\nabla}_t(x,\task)$ is the mean of the posterior belief over the gradient of the constraint induced by our belief over the constraint which, due to properties of GPs, is also a GP. This  is a local version of the approximation proposed in \cite{gonzalez2016batch}. 
Based on this approximation, we want to determine whether, for the current task $\task$, an optimistic observation of the constraint at controller $x$, $l^q_t(x,\task)$ would allow us to classify as safe a controller $z$ despite an $\epsilon$ uncertainty due to noisy observations of the constraint. To this end, we introduce the optimistic noisy expansion operator 
\begin{equation*}
    \g{x}{z}=\mathbb{I}\left[l^q_t(x,\task)+\|\mu^\nabla_t(x,\task)\|_\infty d(x,z)+\epsilon \leq \threshold, \right],
\end{equation*}
where $\mathbb{I}$ is the indicator function and $\threshold$ is the upper limit of the constraint. For a safe $x$, $\g{x}{z}>0$ determines that: (\textit{i}) $z$ can plausibly be safe and (\textit{ii}) evaluating the constraint at $x$ could include $z$ in the safe set.

\looseness=-1
\mypar{Optimization and optimistic safe set}
Normally, \goose{}  explicitly computes the optimistic safe set and uses GP-LCB \cite{srinivas2009gaussian} to determine the next input where to evaluate the objective, $\suggestion$. In other words, \goose{} solves $\suggestion_t=\argmin_{x\in S^o_t}\mu^f_{t-1}(x)-\beta^f_{t-1}\sigma^f_{t-1}(x),$
where $S^o_t$ is the optimistic safe set at iteration $t$. However, this requires a fine discretization of the domain $\domain$ to represent $S^o_t$ as finite set of points in $\domain$, which does not scale to large domains. Moreover, the recursive computation of $S^o_t$ is expensive and not well suited to the fast responses required by adaptive control. Similarly to \cite{Duivenvoorden}, here we rely on particle swarm optimization (PSO) \cite{kiranyaz2014multidimensional} to solve this optimization problem, which checks that the particles belong to the one-step optimistic safe set as the optimization progresses and avoids computing it explicitly. We initialize $m$ particles positioned uniformly at random within the discretized pessimistic safe set with grid resolution $\Delta x$, velocity $\Delta x$ with random sign (\cref{alg:pso_init} of \cref{alg:pso}) and fitness equal to the lower bound of the objective $l^f_t(\cdot)$. If a particle belongs to the optimistic safe set (\cref{alg:pso_optimistic_condition}) and its fitness improves  (\cref{alg:pso_imptovement}), we update its best position. This step lets the particle diffuse into the optimistic safe set without computing it explicitly. Subsequently, we update the particles' positions (\cref{alg:pso_ptilde_update}) and velocities (\cref{alg:pso_v_update}) based on the particles' best position $z_i^j$ and overall best position $\overline{z}^j$, which is updated in \cref{alg:pso_overz_update}.

\looseness=-1
\mypar{Algorithm}
We are ready to present our variant of \goose{} for adaptive control. We start by computing the pessimistic safe set $S_t$ (\cref{alg:Spess} of \cref{alg:goose}) on a grid with lengthscale-dependent resolution $\Delta x$, its boundary $L_t$ (\cref{alg:L}) and the uncertain points on its boundary $W_t$ (\cref{alg:W}), which are used to determine whether controllers belong to the optimistic safe set. Based on these, a new suggestion $\suggestion$ is computed (\cref{alg:get_suggestion}). If its lower bound is close to the best observation for the current task $\xbest(\task)=\argmin_{\{(x',\task',f(x')) \in \data:\task'=\task\}}f(x')$, we stop (\cref{alg:stopping}). Otherwise, if the suggestion is safe, we evaluate it and possibly update the task parameter (\cref{alg:safe_evaluation}). Finally, if we are not sure it is safe, we evaluate all the expanders $x_w^*$ in increasing order of distance from the suggestion $\suggestion$, until either $\suggestion$ is in the pessimistic safe set and can be evaluated or there are no expanders for $\suggestion$ and a new query to PSO is made (\cref{alg:get_expander,alg:evaluate_expander}). During this inner loop the task parameter is constantly updated.

\section{Numerical results}
\label{sec:numerics}

\begin{figure*} [h]
\begin{subfigure}[c]{0.25\textwidth}
\includegraphics[width=1\textwidth]{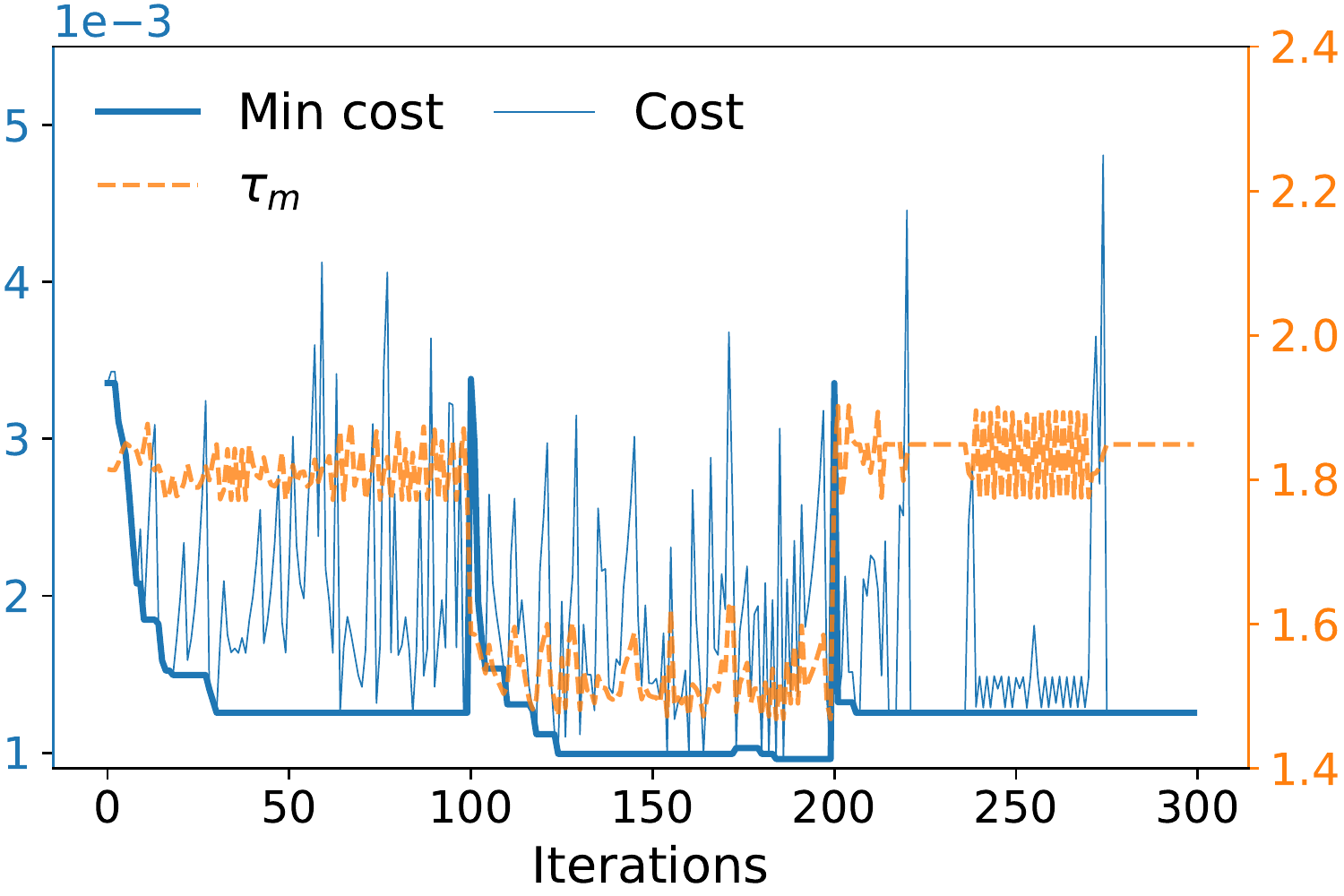}
\label{fig:adaptiveC_sim_m_cost}
\end{subfigure}
\hfill
\begin{subfigure}[c]{0.25\textwidth}
\includegraphics[width=1\textwidth]{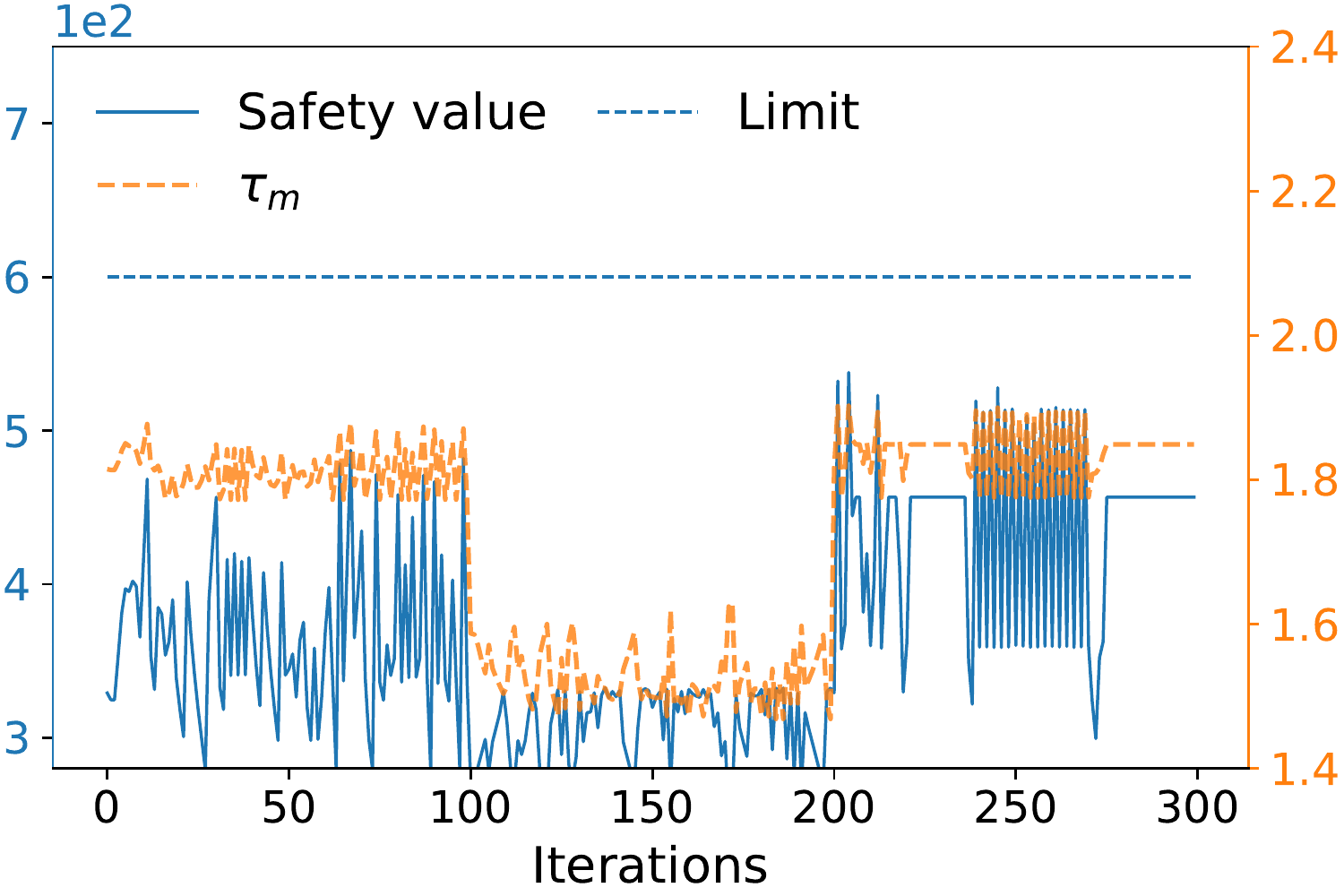}
\label{fig:adaptiveC_sim_m_q1}
\end{subfigure}
\hfill
\begin{subfigure}[c]{0.25\textwidth}
\includegraphics[width=1\textwidth]{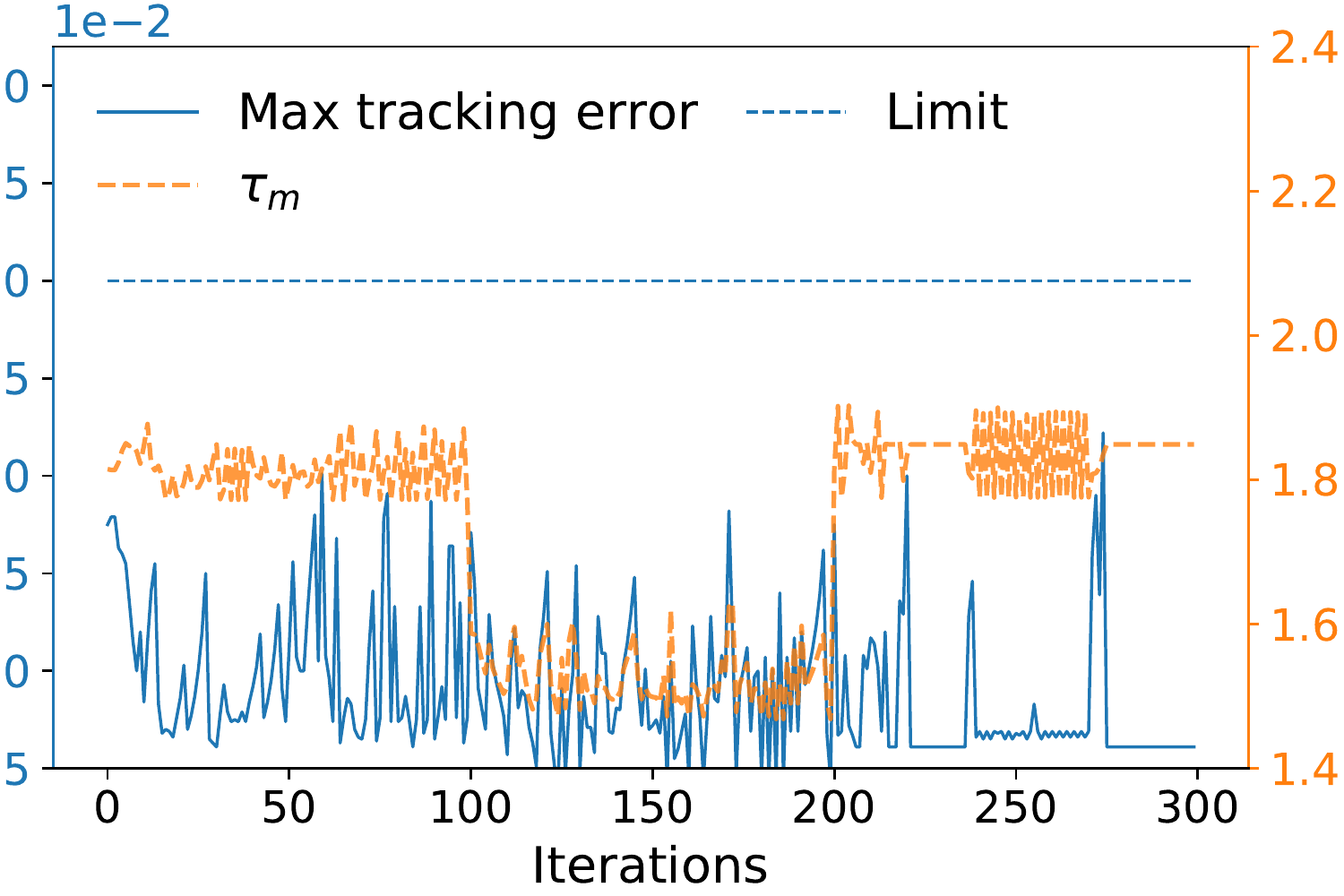}
\label{fig:adaptiveC_sim_m_q2}
\end{subfigure}
\hfill
\footnotesize\addtolength{\tabcolsep}{-4pt}
    \raisebox{0.3cm}{
    \begin{tabular}{ @{}l c c c c@{} }\toprule
    Task ($\task_\text{m}$) & $1.81$ & $1.52$ & $1.84$\\\midrule
    It. to conv. & $100$ & $95$ & $65$\\
    $f(\xbestgrid(\task_\text{m}))$ & $1.27$ & $0.97$ & $1.27$\\
    $f(\xbest(\task_\text{m}))$  & $1.26$ & $0.98$ & $1.26$\\
    $\Kpbest$  & 50 & 49 & 50\\
    $\Kvbest$  & 0.10 & 0.11 & 0.10\\
    $\Tibest$  & 1 & 1 & 1 \\
   \bottomrule
\end{tabular}}
\\[-0.4cm]
\begin{subfigure}[c]{0.26\textwidth}
\includegraphics[width=1\textwidth]{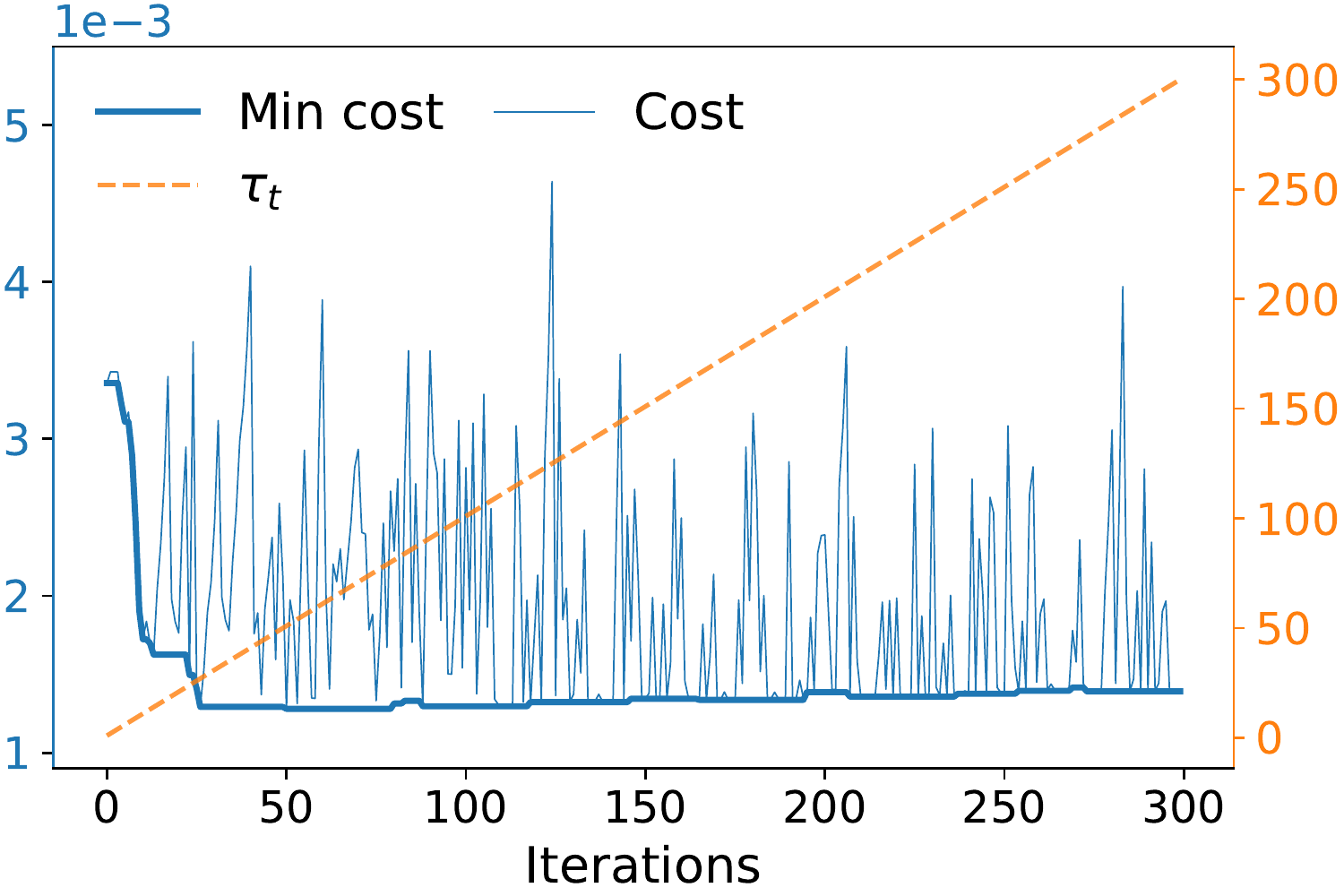}
\caption{Minimum and observed cost $f$}
\end{subfigure} 
\begin{subfigure}[c]{0.26\textwidth}%
\includegraphics[width=1\textwidth]{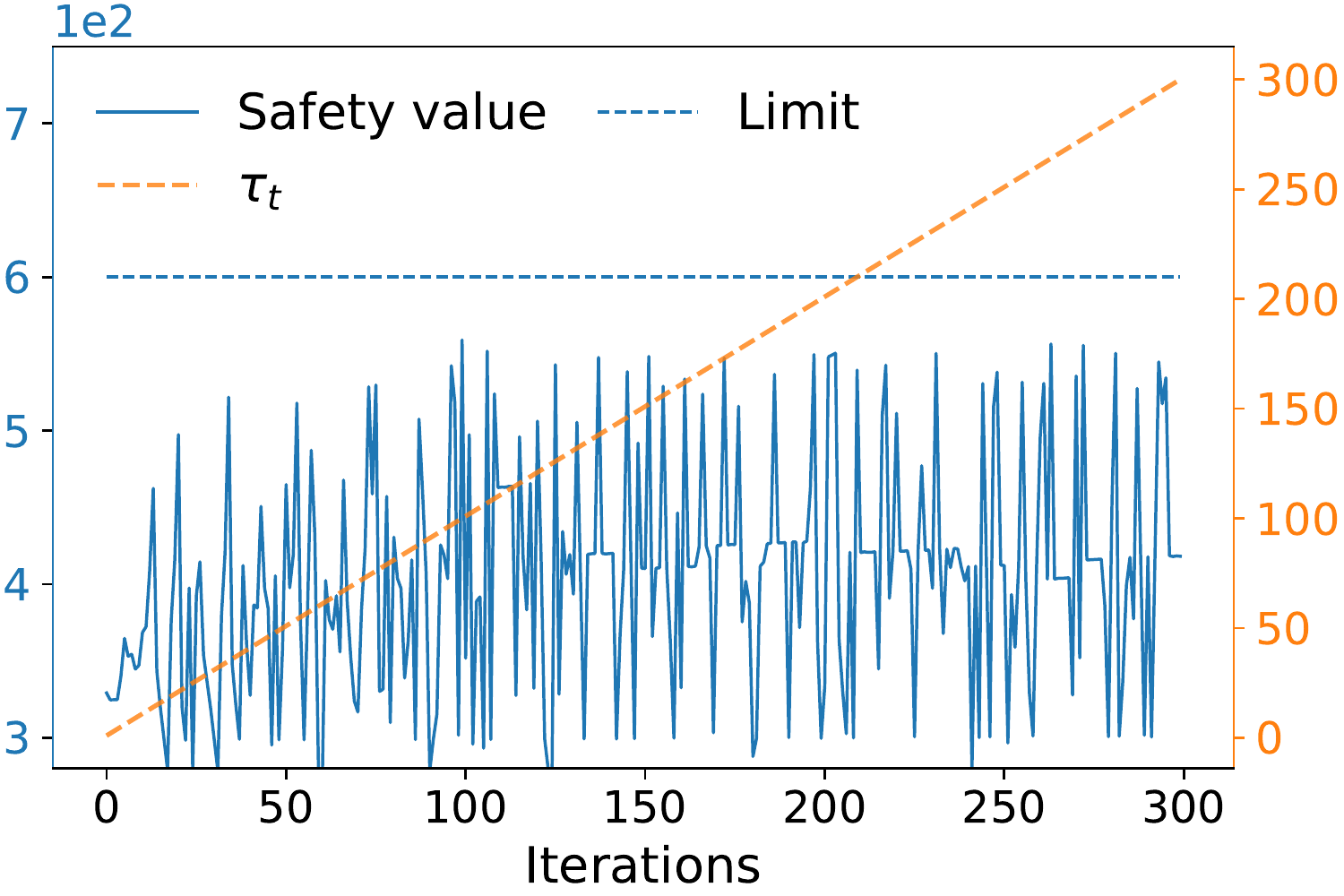} 
\caption{Safety constraint $q_1$.}
\end{subfigure}
\begin{subfigure}[c]{0.26\textwidth}
\includegraphics[width=1\textwidth]{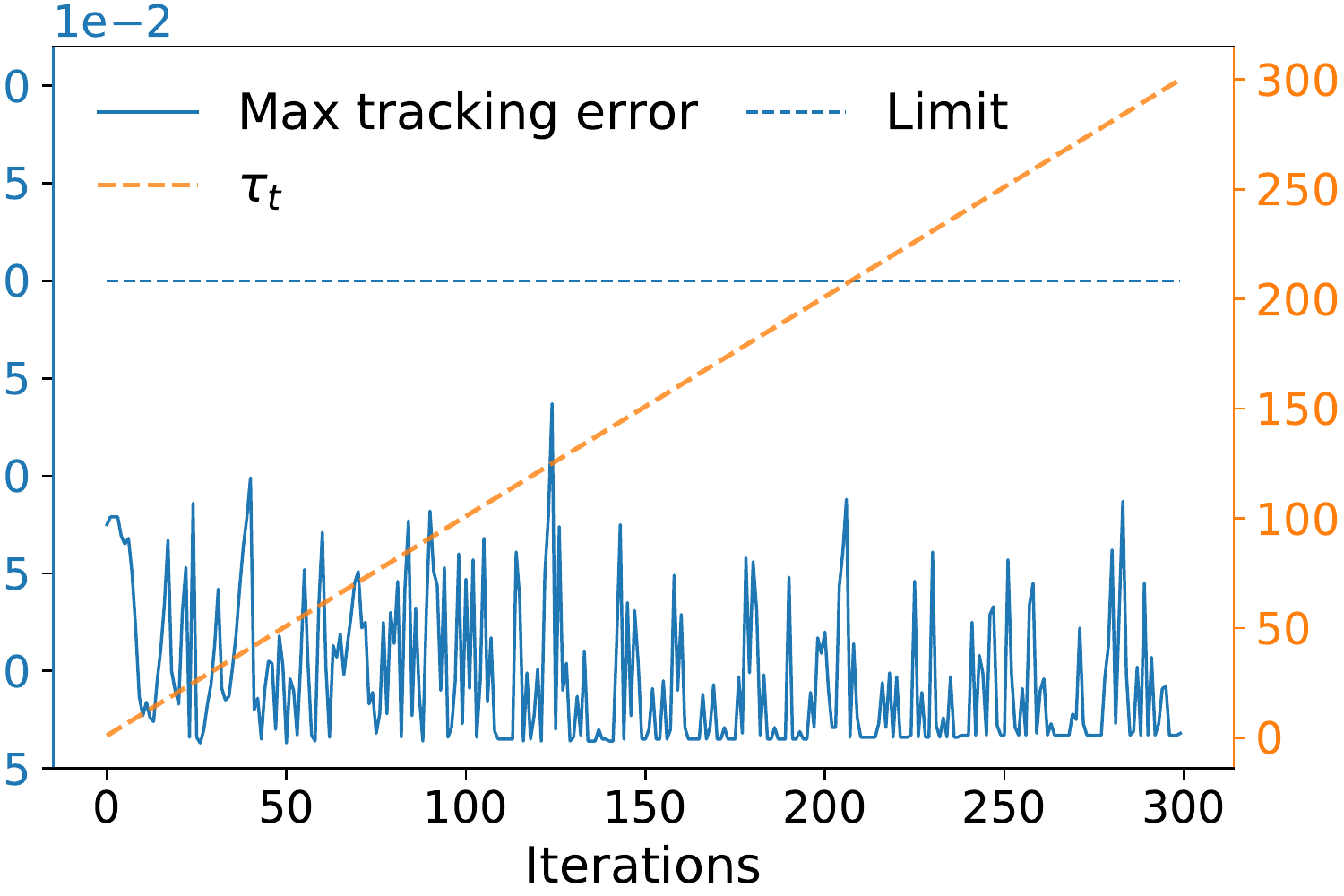}
\caption{Tracking error constraint $q_2$.}
\end{subfigure}
\footnotesize\addtolength{\tabcolsep}{-0pt}
    \raisebox{0.3cm}{
    \begin{tabular}{ @{}l c c c c@{} }\toprule
    Task ($\task_\text{t}$) & $30$ & $300$ \\\midrule
    $f(\xbestgrid(\task_\text{t}))$ & $1.28$ & $1.41$ \\
    $f(\xbest(\task_\text{t}))$  & $1.3$ & $1.39$ \\
    $\Kpbest$  & 46 & 50 \\
    $\Kvbest$  & 0.09 & 0.08 \\
    $\Tibest$  & 1 & 1  \\
   \bottomrule
\end{tabular}}

\caption{\small Cost (left), safety constraint (center) and performance constraint (right) for the simulated adaptive control experiments with sudden change of the moment of inertia $m$ (upper) and a slow change of the rotational damping $b$ (lower). The thins lines show the values for each experiment. The thick one shows the best cost found for the current task. The tables show the mean values over 10 repetitions for these experiments.  \goose{} quickly finds optimal solutions and is able to adapt to both kind of disturbances.}
\vspace{-0.3cm}
\label{fig:adaptiveC_sim}
\end{figure*}

\looseness=-1
We first apply \cref{alg:goose} to tune the controller in \cref{fig:Scheme}, simulating the system model in \cref{sec:problem_statement} in stationary conditions. Later, we use our method for adaptive control of instantaneous and slow-varying changes of the plant. In \cref{sec:appendix}, we present an ablation study that investigates the impact of the task parameter on these problems.

\looseness=-1
The optimization ranges of the controller parameters are set to $\Kp \in [5,50]$, $\Kv \in [0.1, 0.11]$ for the position and velocity gains, and $\Ti \in [1,10]$ for the time constant. For each task, \goose{} returns the controller corresponding to the best observation,  $\xbest(\task)=(\Kpbest,\Kvbest,\Tibest)$. The cost $f(\mathrm{x})$ is provided in $[\deg10^{-3}]$ units. We denote the `true` optimal controller computed via grid search as $\xbestgrid$. Due to noise, not even this controller can achieve zero tracking error, see the table in \cref{fig:cbo_comparison}. We use a zero mean prior and squared exponential kernel with automatic relevance determination, with  length scales for each dimension $[l_{\Kp}, l_{\Kv}, l_{\Ti}, l_\text{m},l_{K_{\text{ff}}},l_{\tilde{b}}]=[30,0.03,3,0.5,0.3,5]$, identical for $f$, $q_1$ and $q_2$ for the numerical simulations and the experiments on the system. The likelihood variance is adjusted separately for each GP model. 

\looseness=-1
\mypar{Control for stationary conditions} 
In this section, we compare \goose{} to CBO for controller tuning \cite{Koenig2020} in terms of the cost and the safety and performance constraints introduced in \cref{sec:problem_statement} when tuning the plant under stationary conditions. We benchmark these methods against exhaustive grid-based evaluation using a grid with $5\times11\times10$ points.

For each algorithm, we run 10 independent experiments, which vary due to noise injected in the simulation. Each experiment starts from the safe controller $[\Kp,\Kv,\Ti] = [15,0.05,3]$. The table in \cref{fig:cbo_comparison} shows the median number of iterations needed to minimize the cost. \cref{fig:cbo_comparison} (left) shows the convergence of both algorithms for each repetition. While both algorithms converge to the optimum, \goose{} prevents constraint violations for all iterations (see \cref{fig:cbo_comparison}). In contrast,  CBO violates the constraints in $27.8\%$ of the iterations to convergence, reaching far in the unsafe range beyond the safety limit of acceptable vibrations in the system (\cref{fig:cbo_comparison}, center), showing that additional safety-related measures are required. While the constraint violations incurred by CBO can be limited in stationary conditions by restricting the optimization range, this is not possible for adaptive control. 

\looseness=-1
\mypar{Adaptive control for instantaneous changes}
We show how our method adapts to an instantaneous change in the load of the system. We modify the moment of inertia $m$ and estimate this from the system data. We inform the algorithm of the operating conditions through a task parameter, $\task_\text{m} = \log10(\frac{1}{N}\sum_{1}^N|\text{FFT}(v - v_\text{ff})|)$, which is calculated from the velocity measurement $v$,  using the feed-forward signal $v_\text{ff}$ from position to velocity. This data-driven task parameter exploits the differences in the levels of noise in the velocity signal corresponding to different values of the moment of inertia $m$. As the velocity also depends on the controller parameters, configurations in a range of $\pm 0.15$ of the current $\task_m$ value are treated as the same task configuration. The algorithm is initialized with $[\Kp, \Kv, \Ti] = [15,0.05,3]$ as safe seed for all tasks. The moment of inertia of the plant $m$  is
switched every $100$ iterations. The table in \cref{fig:adaptiveC_sim} summarizes 10 repetitions of the experiment with a stopping criterion set to $\epsilon_\text{tol} = 0.002$.
\begin{figure*}[t]
\raisebox{0.2cm}{
\begin{subfigure}[c]{0.23\textwidth}
\label{fig:adaptiveC_exp_Kff_cost}
\includegraphics[width=1\textwidth]{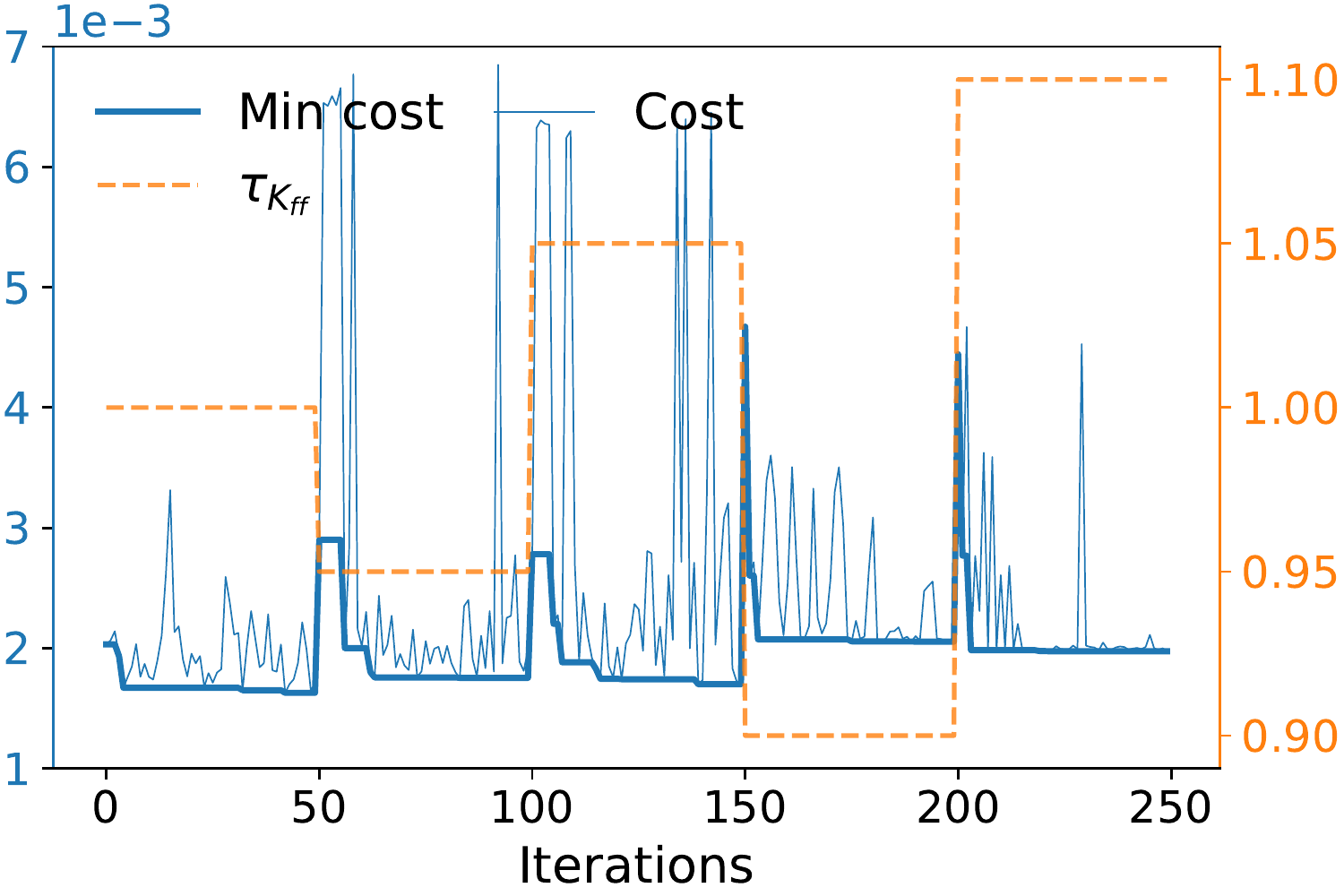}
\end{subfigure}}
\hfill
\begin{subfigure}[c]{0.23\textwidth}
\includegraphics[width=1\textwidth]{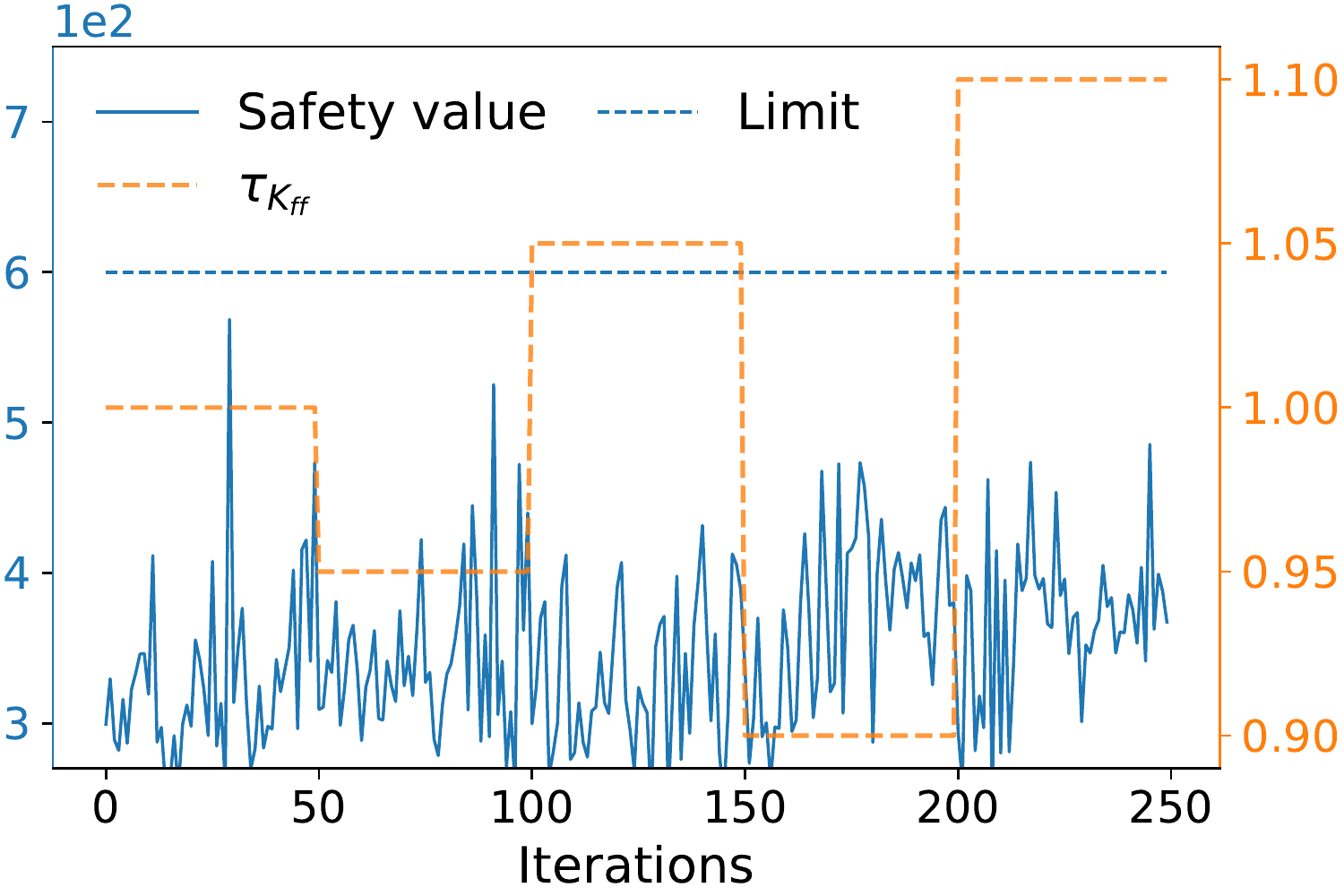}
\label{fig:adaptiveC_exp_Kff_q1}
\end{subfigure}
\hfill
\begin{subfigure}[c]{0.23\textwidth}
\includegraphics[width=1\textwidth]{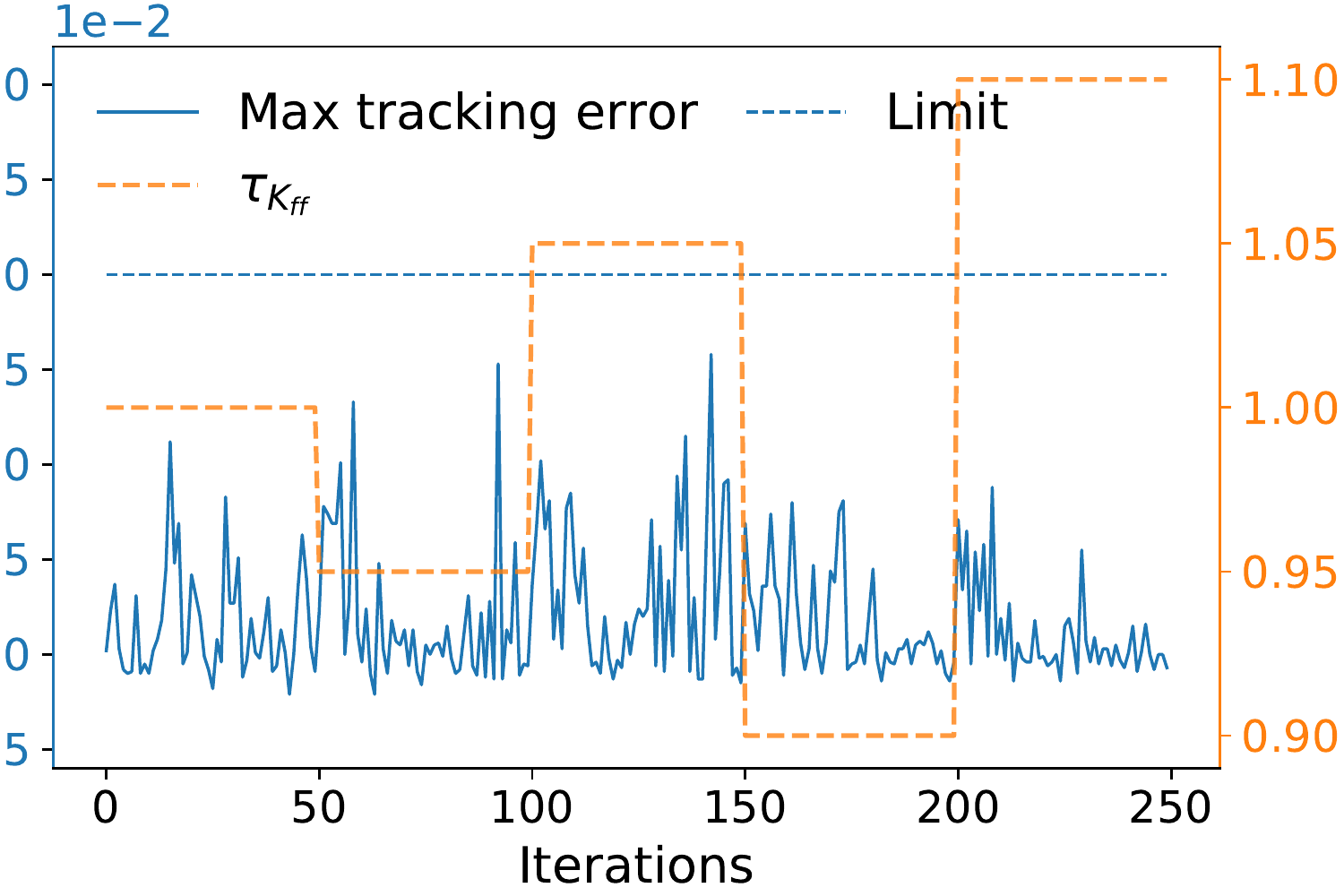}
\label{fig:adaptiveC_exp_Kff_q2}
\end{subfigure}
\label{fig:adaptiveC_exp_Kff}
\hfill
\footnotesize\addtolength{\tabcolsep}{-5pt}
\raisebox{0.3cm}{
\begin{tabular}{ @{}l c c c c c c@{} }\toprule
Task ($\task_{K_{ff}}$) & $1.0$ & $0.95$ & $1.05$ & $0.9$ & $1.1$\\ \midrule
It. to conv. & $50$ & $49$ & $47$ & $33$ & $16$\\
$f(\xbest(\task_{K_{ff}}))$ & $1.63$ & $1.75$ & $1.7$ & $2.07$ & $1.98$\\
$\Kpbest$ & $37$ & $43$ & $50$ & $50$ & $50$\\
$\Kvbest$ & $.08$ & $.09$ & $.09$ & $.09$ & $.09$\\
$\Tibest$ & $1$ & $10$ & $10$ & $9$ & $8$\\
\bottomrule
\end{tabular}}
\\[-0.5cm] 
\begin{subfigure}[c]{0.25\textwidth}
\includegraphics[width=1\textwidth]{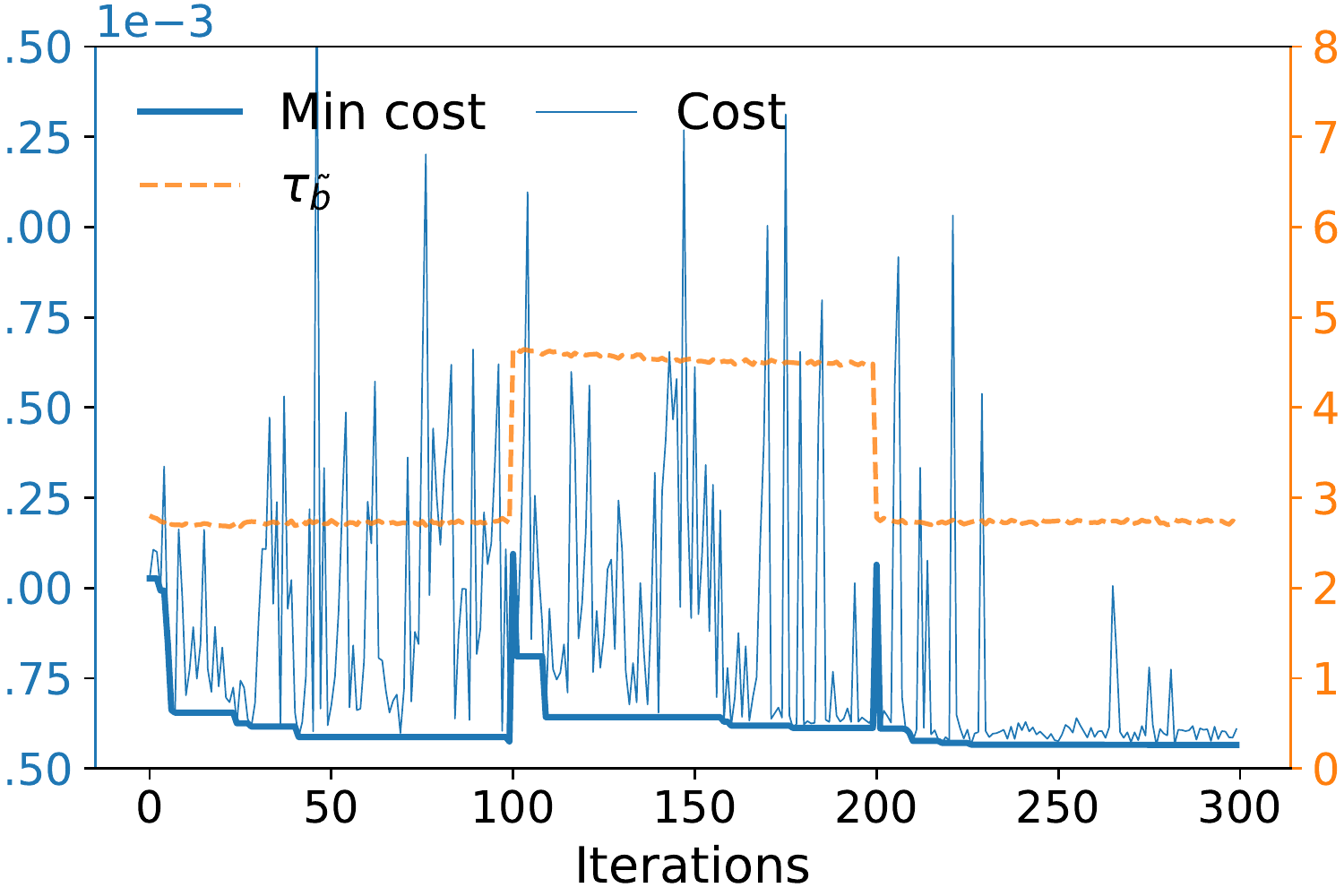}
\caption{Minimum and observed cost $f$.}
\label{fig:adaptiveC_exp_b_cost}
\end{subfigure}
\hfill
\begin{subfigure}[c]{0.25\textwidth}
\includegraphics[width=1\textwidth]{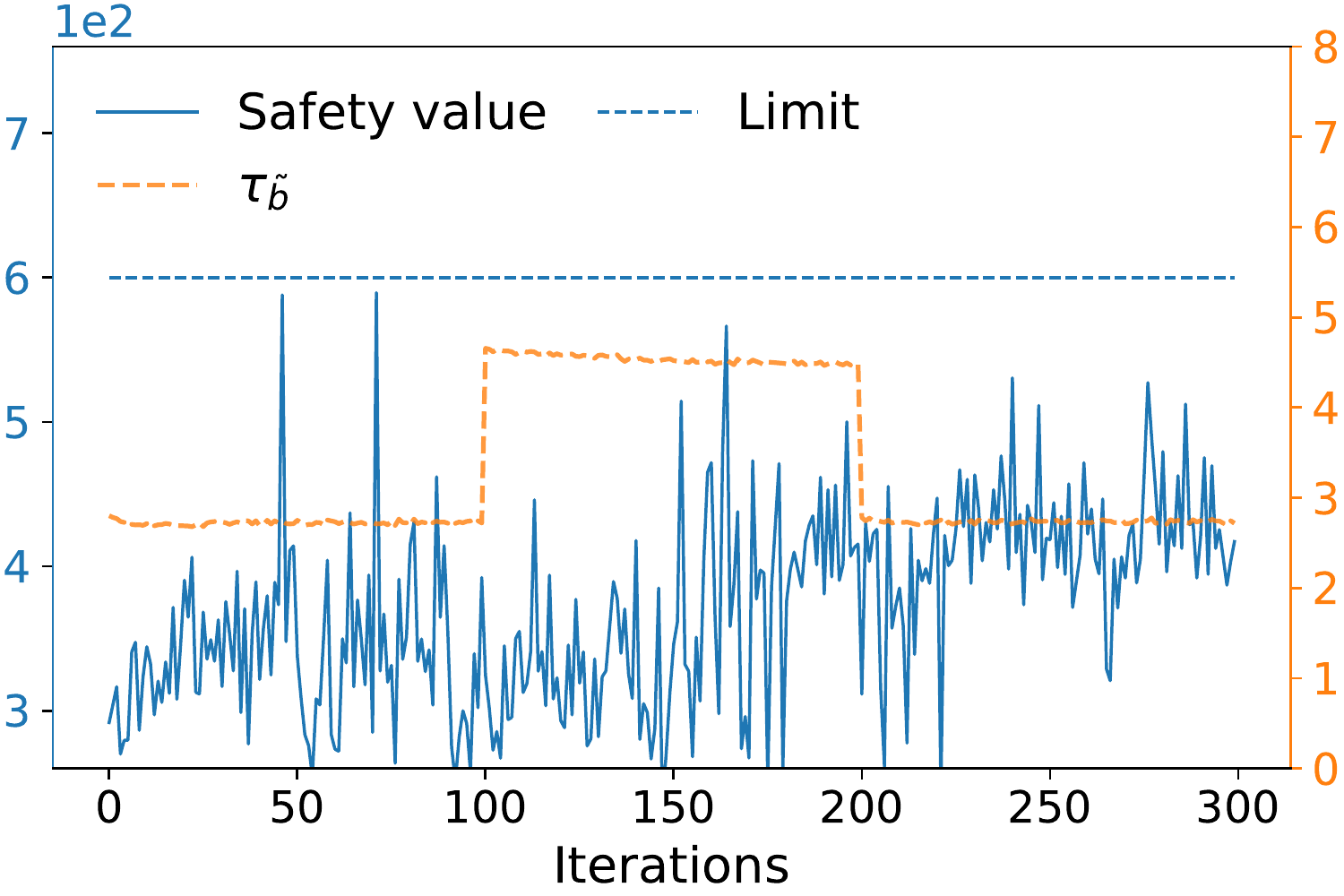}
\caption{Safety constraint $q_1$.}
\label{fig:adaptiveC_exp_b_q1}
\end{subfigure}
\hfill
\begin{subfigure}[c]{0.25\textwidth}
\includegraphics[width=1\textwidth]{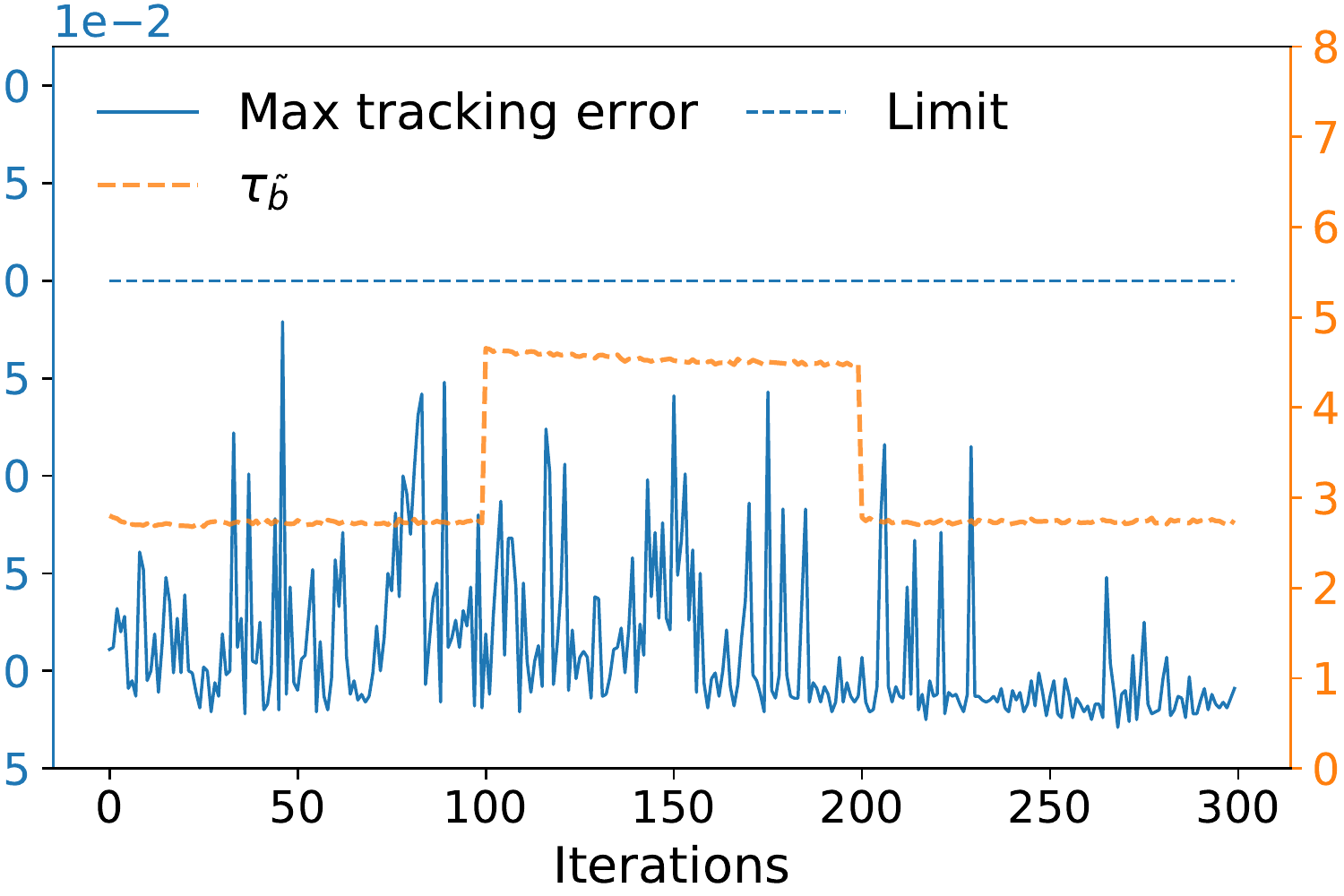}
\caption{Tracking error constraint $q_2$.}
\label{fig:adaptiveC_exp_b_q2}
\end{subfigure}
\hfill
\raisebox{0.3cm}{
\footnotesize\addtolength{\tabcolsep}{-0pt}
    \begin{tabular}{ @{}l c c c c@{} }\toprule
    Task ($\task_{\tilde{b}}$) & $2.72$ & $4.53$ & $2.73$\\\midrule
    It. to conv. & $99$ & $78$ & $19$\\
    $f(\xbest(\task_{\tilde{b}}))$ & $1.59$ & $1.61$ & $1.59$\\
    $\Kpbest$ & $32$ & $30$ & $37$\\
    $\Kvbest$ & $.09$ & $.09$ & $.1$\\
    $\Tibest$ & $6$ & $7$ & $7$\\
   \bottomrule
\end{tabular}}
\caption{\looseness=-1\small Cost (left), safety constraint (center), performance constraint (right) and summary (table) for the real-world adaptive control experiments with sudden change of the feed-forward gain $K_{\text{ff}}$ (upper) and rotational damping $\tilde{b}$ (lower). The thins lines show the values for each experiment. The thick one shows the best cost found for the current task. \goose{} quickly finds optimal controllers for all the regimes (left) without violating the constraints (center-right) despite the location of the optimum keeps changing (table).}
\label{fig:adaptiveC_exp_b}
\vspace{-0.3cm}
\end{figure*}
\cref{fig:adaptiveC_sim} (top left) shows the sampled cost and the task parameters for one run, and the running minimum for each task. We clearly see that \goose{} quickly finds a high performing controller for the initial operating conditions ($m$ = 0.0191$\mathrm{kgm^2}$). Subsequently, it adapts $\Kp$ and $\Kv$ in few iterations to the new regime ($m$ = 0.0382$\mathrm{kgm^2}$) due to the capability of the GP model to generalize across tasks (\cref{fig:adaptiveC_sim}, top table). Finally, when the system goes back to the initial regime, \goose{} immediately finds a high-performing controller and manages to marginally improve over it. In particular, the best cost $f(\xbest(\task_\text{m}))$ found for each condition coincides with the optimum obtained by grid evaluation $f(\xbestgrid(\task_\text{m}))$. Moreover, \cref{fig:adaptiveC_sim} (top center left and right) shows that the constraints are never violated. 

\mypar{Adaptive control for gradual changes}
\looseness=-1
We show how our method adapts to slow changes in the dynamics. In particular, we let the rotation damping coefficient increase linearly with time: $b(t) = b_0(1+\frac{t}{1000})$, where $b_0=30.08\mathrm{kgm^2/s}$. In this case, time is the task parameter and, therefore, we use the temporal kernel $k(t,t') = (1 - \epsilon_t)^{|t - t'|/2}$ \cite{bogunovic2016timevarying} as task kernel with $\epsilon_t = 0.0001$. This kernel increases the uncertainty in already evaluated samples with time. To incorporate the drift of the dynamics, we evaluate the best sample $\xbest$ in Algorithm \ref{alg:goose} with respect to the stopping criterion threshold $\epsilon_\text{tol} = 0.001$ in a moving window of the last $30$ iterations. Learning is slower, due to the increased uncertainty of old data points. On average, the stopping criterion (\cref{alg:stopping} in \cref{alg:goose}) is fulfilled in $74$ out of $300$ iterations. The optimum increases, (\cref{fig:adaptiveC_sim}, bottom left and table), corresponding to a change in the controller parameters. In the second half of the experiment the cost decreases, and reaches optimum more often, as the GPs successfully learn the trend of the  parameter $b(t)$. There were no constraint violations in any of the repetitions, see \cref{fig:adaptiveC_sim} (bottom center left and right).

\section{Experimental results}
\label{sec:experimental}

\looseness=-1
We now demonstrate the proposed adaptive control algorithm on a rotational drive system. The system has an encoder resolution of $\Delta p = 1\deg\times10^{-7}$ for the angular position, $\Delta v = 0.0004~\text{RPM}$ for the angular velocity and $\Delta T = 0.0008$~Nm for the torque. The angular position of the system has no hardware limit. The limits of the angular velocity and the torque are $v_\text{lim} = 50~\text{RPM}$ and $T_\text{lim} = 3.48$~Nm, respectively. 
First, we show how the controller parameters adapt when the algorithm is explicitly informed about a change in the feed-forward gain $K_\text{ff}$. We then demonstrate the performance when an external change occurs, corresponding to change in the rotational resistance, which can be estimated from the system's data. The optimization ranges and the kernels hyperparameters are the same as in \cref{sec:numerics}. The likelihood variance is adjusted separately for each GP model.

\mypar{Adaptive control for internal parameter change} 
We start an experiment with the nominal feed-forward gain, $K_{\mathrm{ff}}=1$, and switch it subsequently four times between $0.9$ and $1.1$ in intervals of 50 iterations. For each value of $K_{\mathrm{ff}}$, the starting point is a safe sample, collected with $[\Kp, \Kv, \Ti] = [15,0.05,3]$. The value of $K_{\mathrm{ff}}$ is used as task parameter $\task_{K_{\text{ff}}}$. The stopping criterion is set to $\epsilon_\text{tol} = 0.001$. 

The convergence of the optimization accelerates with increasing data. \cref{fig:adaptiveC_exp_b} (top left) shows that convergence is not reached for the nominal $K_{\text{ff}}$ during the first $50$ iterations, whereas the last configuration with $\task_{K_{\text{ff}}}=1.1$ requires only $16$ until convergence, showing that learning is efficient, even if the optimum shifts w.r.t. the configuration. Constraint violations are completely prevented for all tasks parameters, as shown in \cref{fig:adaptiveC_exp_b} (top center and right).

\looseness=-1
\mypar{Adaptive control for external parameter change}
We now validate experimentally the change introduced to the controller by modifying the friction $\tilde{b}$ in the system, which is related to the non-linearity in the dynamics of the plant. We estimate the change in $\tilde{b}$ by the average torque measurement of the system and provide it as task parameter $\task_{\tilde{b}}$, as shown in \cref{fig:adaptiveC_exp_b} (bottom). We start with optimizing the nominal controller parameters for 100 iterations, followed by an increase of $\tilde{b}$ (and of $\task_{\tilde{b}}$ accordingly), which is achieved by wrapping elastic bands around the rotational axis and fixing them at the frame of the system. In the last 100 iterations we switch back to the nominal condition. The stopping criterion is set to $\epsilon_\text{tol} = 0.001$. Since $\task_{\tilde{b}}$ is not fixed and is influenced by noise, all configurations with $\task_{\tilde{b}}$ in a range of $\pm 0.1$ of the current $\task_{\tilde{b}}$ value are treated as the same task configuration. The cost increases after the first intervention, then reaches close to nominal values in 10 iterations, and adapts almost instantaneously at the next $\task_{\tilde{b}}$ switch (\cref{fig:adaptiveC_exp_b}, bottom left). The constraints are never violated (\cref{fig:adaptiveC_exp_b}, bottom center,  right).

\section{Conclusion}
\looseness=-1
We present a model-free approach to safe adaptive control.
To this end, we introduce several modifications to \goose{}, a safe Bayesian optimization method, to enable its practical use on a rotational motion system.
We demonstrate numerically and experimentally that our approach is sample efficient, safe, and achieves the optimal performance for different types of disturbances encountered in practice. Our approach can be further extended by including multiple performance metrics in the optimization objective, or gradually tightening the constraints, once sufficient system data is available. 

\section{Appendix}
\begin{figure*} [!htb]
\begin{subfigure}[c]{0.24\textwidth}
\includegraphics[width=1\textwidth]{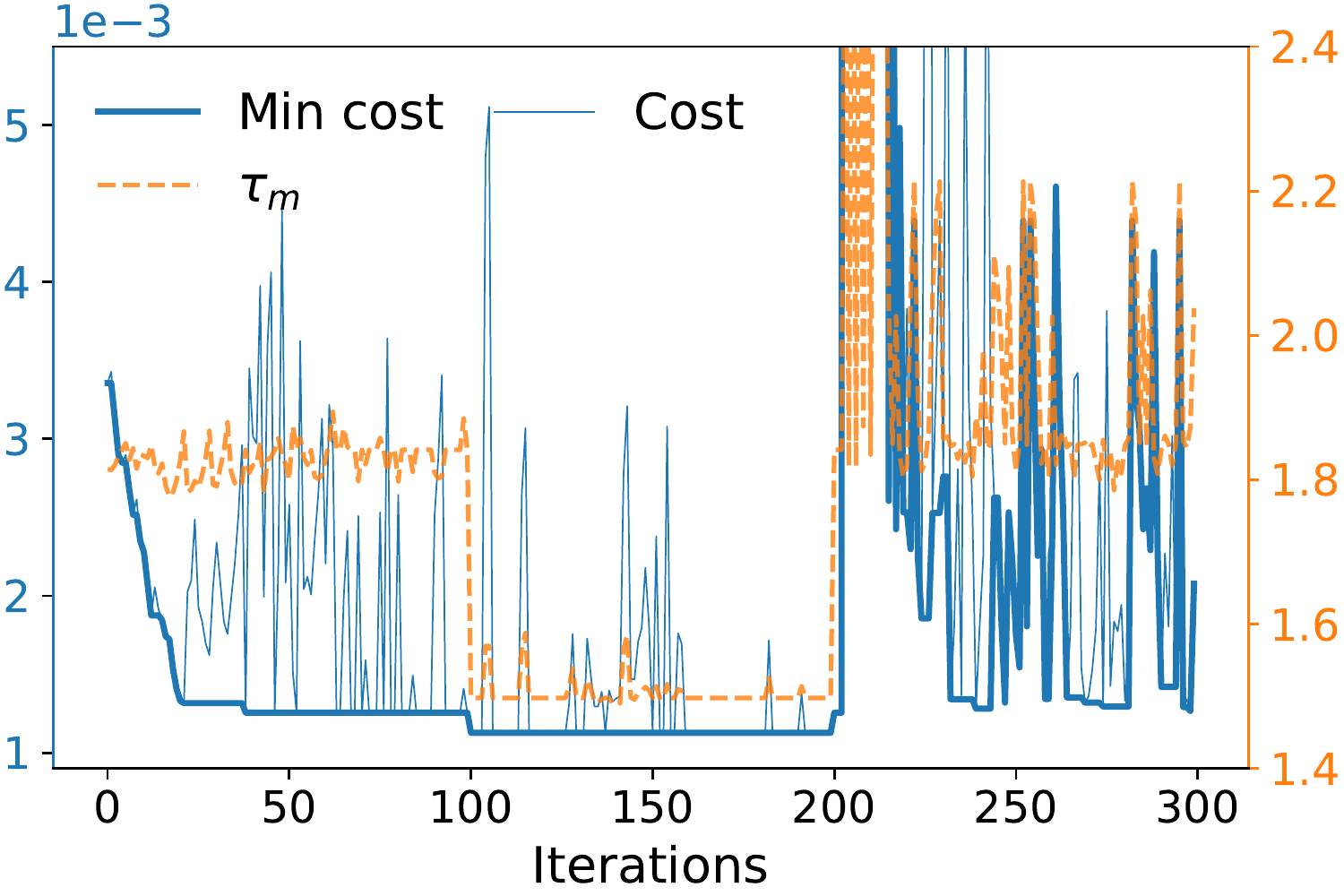}
\label{fig:adaptiveC_sim_m_no_task_cost}
\end{subfigure}
\hfill
\begin{subfigure}[c]{0.24\textwidth}
\includegraphics[width=1\textwidth]{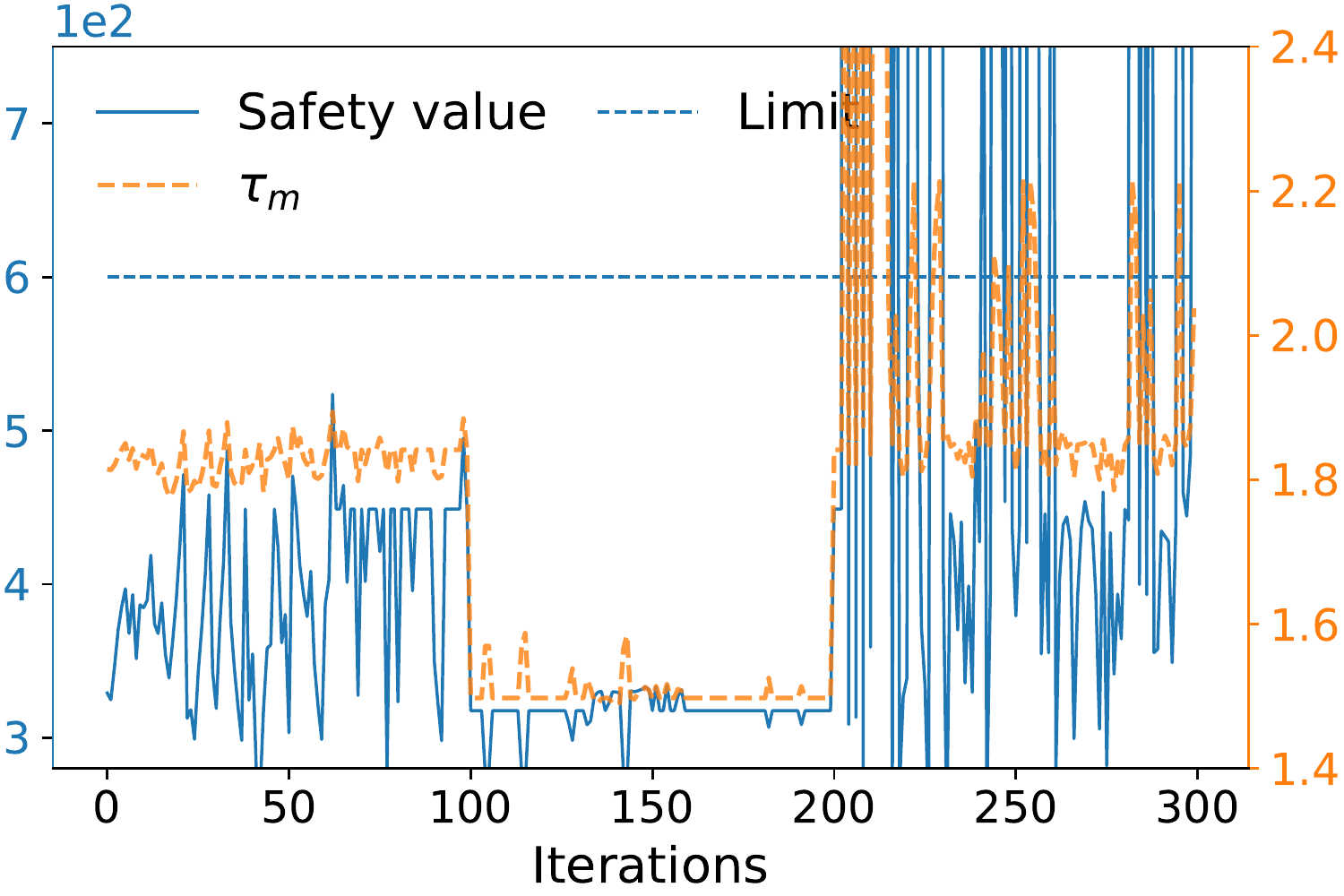}
\label{fig:adaptiveC_sim_m_no_task_q1}
\end{subfigure}
\hfill
\begin{subfigure}[c]{0.24\textwidth}
\includegraphics[width=1\textwidth]{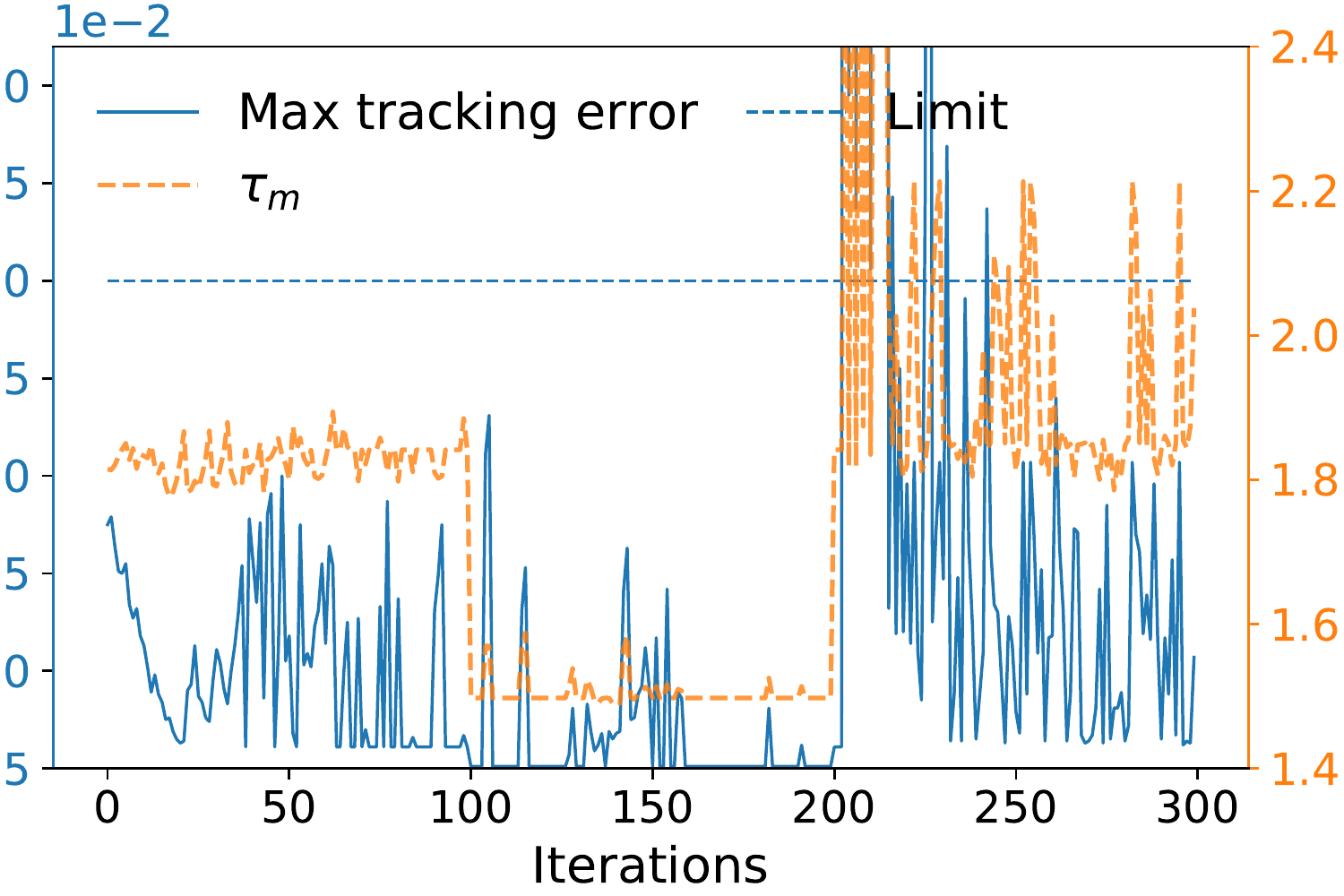}
\label{fig:adaptiveC_sim_m_no_task_q2}
\end{subfigure}
\hfill
\footnotesize\addtolength{\tabcolsep}{-4pt}
    \raisebox{0.3cm}{
    \begin{tabular}{ @{}l c c c c@{} }\toprule
    $\task_\text{m}$ & $1.83$ & $1.5$ & $2$\\\midrule
    $f(\xbestgrid(\task_\text{m}))$ & $1.27$ & $0.97$ & $1.27$\\
    $f(\xbest(\task_\text{m}))$  & $1.26$ & $1.03$ & $1.39$\\
    $\Kpbest$  & 50 & 46.73 & 46.73\\
    $\Kvbest$  & 0.095 & 0.11 & 0.11\\
    $\Tibest$  & 1 & 1 & 1 \\
   \bottomrule
\end{tabular}}
\\[-0.4cm]
\begin{subfigure}[c]{0.25\textwidth}
\includegraphics[width=1\textwidth]{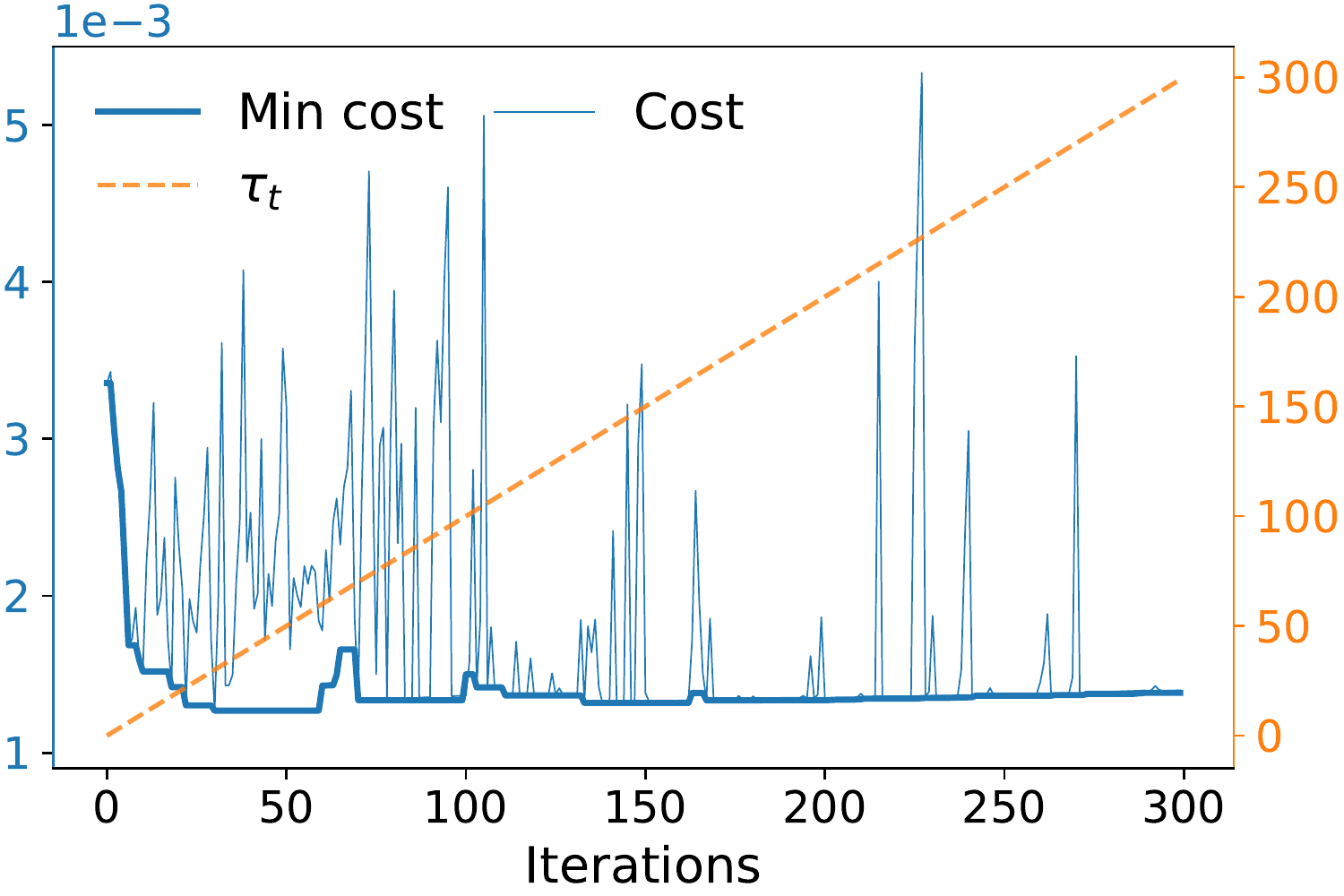}
\caption{Minimum and observed cost $f$}
\end{subfigure} 
\begin{subfigure}[c]{0.25\textwidth}%
\includegraphics[width=1\textwidth]{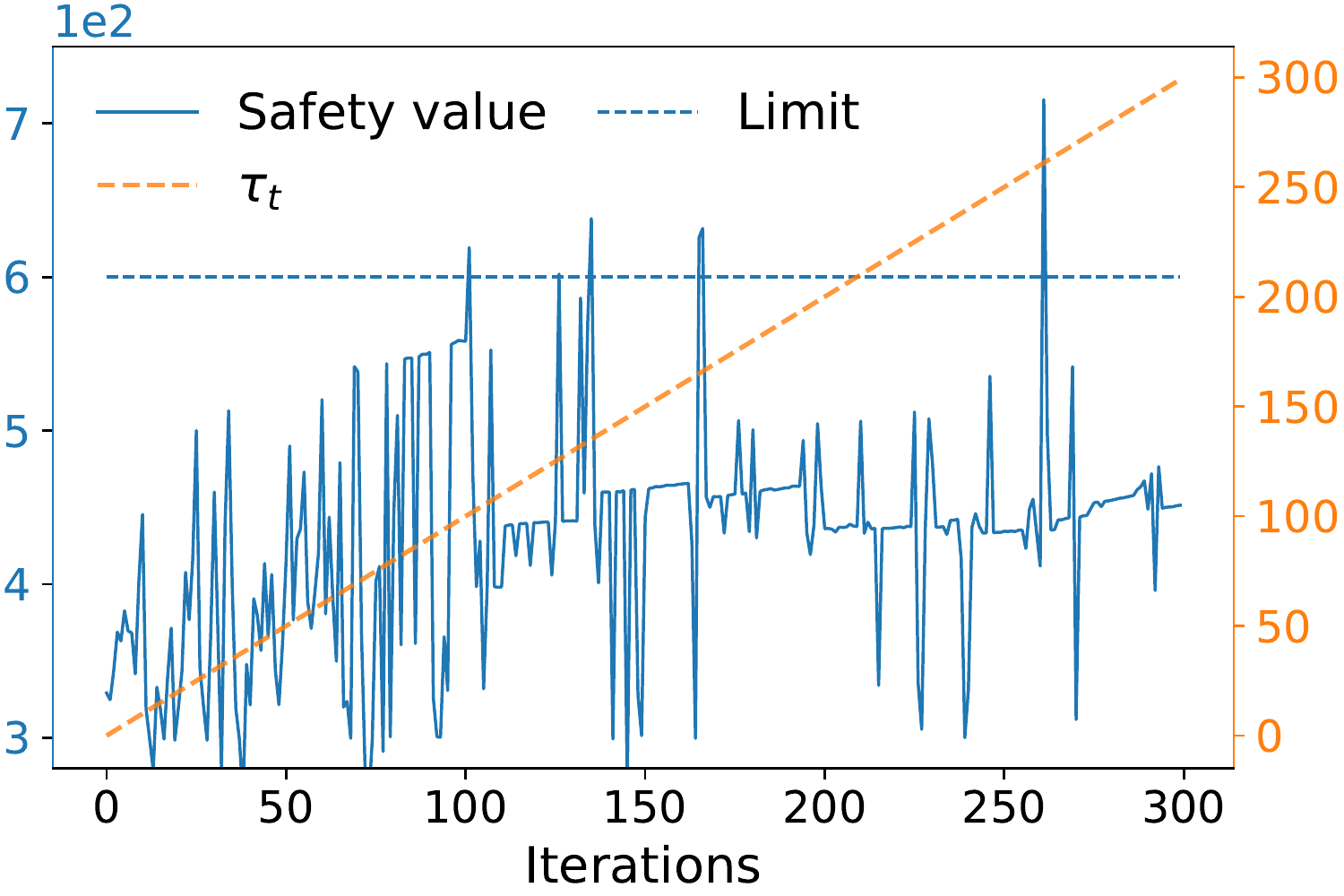} 
\caption{Safety constraint $q_1$.}
\end{subfigure}
\begin{subfigure}[c]{0.25\textwidth}
\includegraphics[width=1\textwidth]{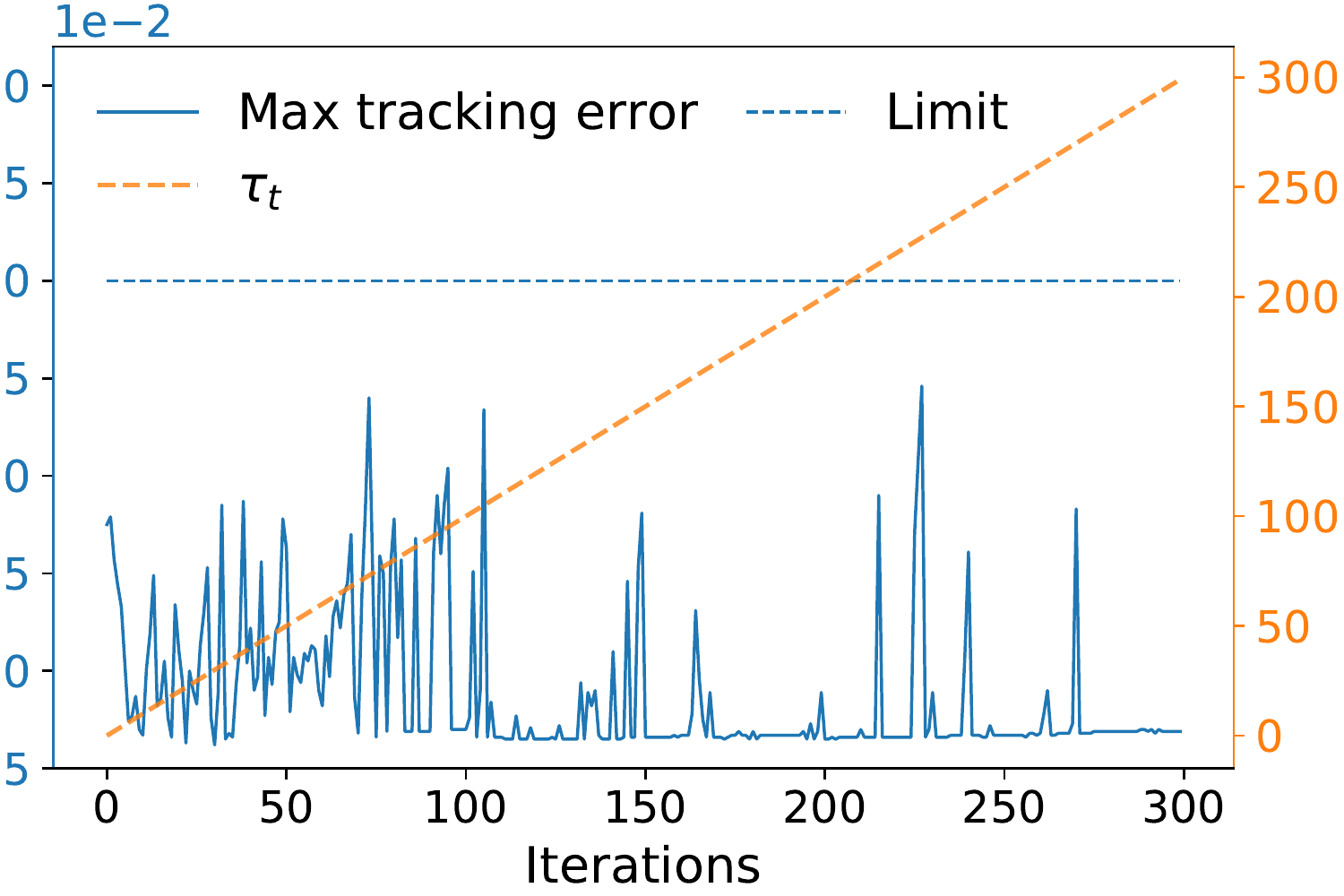}
\caption{Tracking error constraint $q_2$.}
\end{subfigure}
\footnotesize\addtolength{\tabcolsep}{-0pt}
    \raisebox{0.3cm}{
    \begin{tabular}{ @{}l c c c c@{} }\toprule
    $\task_\text{t}$ & $30$ & $300$ \\\midrule
    $f(\xbestgrid(\task_\text{t}))$ & $1.28$ & $1.41$ \\
    $f(\xbest(\task_\text{t}))$  & $1.30$ & $1.40$ \\
    $\Kpbest$  & 46.7 & 50 \\
    $\Kvbest$  & 0.095 & 0.086 \\
    $\Tibest$  & 1 & 1  \\
   \bottomrule
\end{tabular}}
\caption{\small Cost (left), safety constraint (center) and performance constraint (right) for the simulated optimization experiments with sudden change of the moment of inertia $m$ (upper) and a slow change of the rotational damping $b$ (lower) without tracking of the task parameter. The tables summarize the optimization experiments for 10 repetitions. Without task parameter tracking \goose{} results in unsafe behavior for both kind of disturbances.}
\vspace{-0.3cm}
\label{fig:adaptiveC_sim_no_task}
\end{figure*}
\label{sec:appendix}
Here, we study the impact of tracking the task parameter and adapting to its changes on the performance and safety of the algorithm. To this end, we repeat the numerical experiments for instantaneous and gradual changes presented in \cref{sec:numerics} and the experiment for internal parameter changes on the real system presented in  \cref{sec:experimental} without explicitly modeling the task parameter in our algorithm, and compare the results with those obtained modeling it. 

In general, the change in systems dynamics that are induced by a change in external conditions and that are captured by the task parameter greatly impacts the performance and safety of the system when a given controller is applied.  As a result, for a given controller $x$, the algorithm is likely to observe widely different cost and constraints values depending on the task. Without the additional degree of freedom provided by the explicit modelling of the task parameter, the GP model must do its best to make sense out of widely different observations corresponding to the same controller that are too large to be explained by the observation noise. Therefore, we expect this degradation in the quality of the model to induce poor predictive distributions and, therefore, to lead to undesirable and unpredictable behavior of the algorithm.

\mypar{Numerical results}
We start by showing the results obtained for the numerical experiments in  \cref{fig:adaptiveC_sim_no_task}. The top panel shows the performance of the algorithm for the instantaneous change of the moment of inertia in the system, and the bottom panel demonstrates the performance during a gradual linear change in the rotational damping. In both cases, the algorithm does not model these changes. 

The first experiment replicates exactly the one we presented in the top of \cref{fig:adaptiveC_sim}. Initially, $m$ is $0.0191\mathrm{kgm^2}$, which induces an average value of $\task_m$ over the duration of the task of $1.83$, then doubles to $0.0382 \mathrm{kgm^2}$, which induces a lower average value $\task_m=1.5$ as the higher moment of inertia reduces the effect of the noise, and switches back to $0.0191 \mathrm{kgm^2}$, which, due to the model mismatch, induces an even higher average value of $\task_m=2$. Notice that, even though the task parameter $\task_m$ is not provided to the algorithm, we still plot it to show where the tasks switch. During the initial phase of the optimization, the algorithm only experiences one task and, therefore, it is not affected by the fact that it is not modeling the task explicitly. In this phase,  an optimum is found after the initial $100$ iterations. In the second phase, when $m$ increases, the control task becomes effectively easier. Therefore, the algorithm still manages to control the system and to slightly improve the performance. In this phase, aggressive controllers are classified as safe. This is because they do not induce dangerous vibrations in the system due to the increased moment of inertia. However, switching back to the first, lower value of $m$ creates problems. This is due to two main facts: (\textit{i}) controllers that were previously deemed safe due to the high value of $m$ are not anymore. However, since $\task_m$ is not modelled, the algorithm is not aware of this. (\textit{ii}) the GP is conditioned on highly conflicting observations induced by the different tasks that cannot explain through measurement noise. As a consequence, the confidence intervals it provides deteriorate. This results in multiple violations of both constraints.
The second numerical experiment replicates the one shown at the bottom of \cref{fig:adaptiveC_sim}. 
Although the task parameter is no longer tracked, the 30 iterations window for $\xbest$ in \cref{alg:goose} is kept in the algorithm as an adaptive component. However, this adaptive component alone is not sufficient since the GPs are not able to distinguish between old, less accurate data and new, more accurate data without keeping track of the task parameter. Therefore, as the rotation damping coefficient $b$ increases, the system becomes less robust towards aggressive controllers and the algorithm starts using unsafe parameter settings that used to be safe in previous iterations (see figure \cref{fig:adaptiveC_sim_no_task} bottom).
\begin{figure*}[t]
\raisebox{0.2cm}{
\begin{subfigure}[c]{0.25\textwidth}
\includegraphics[width=1\textwidth]{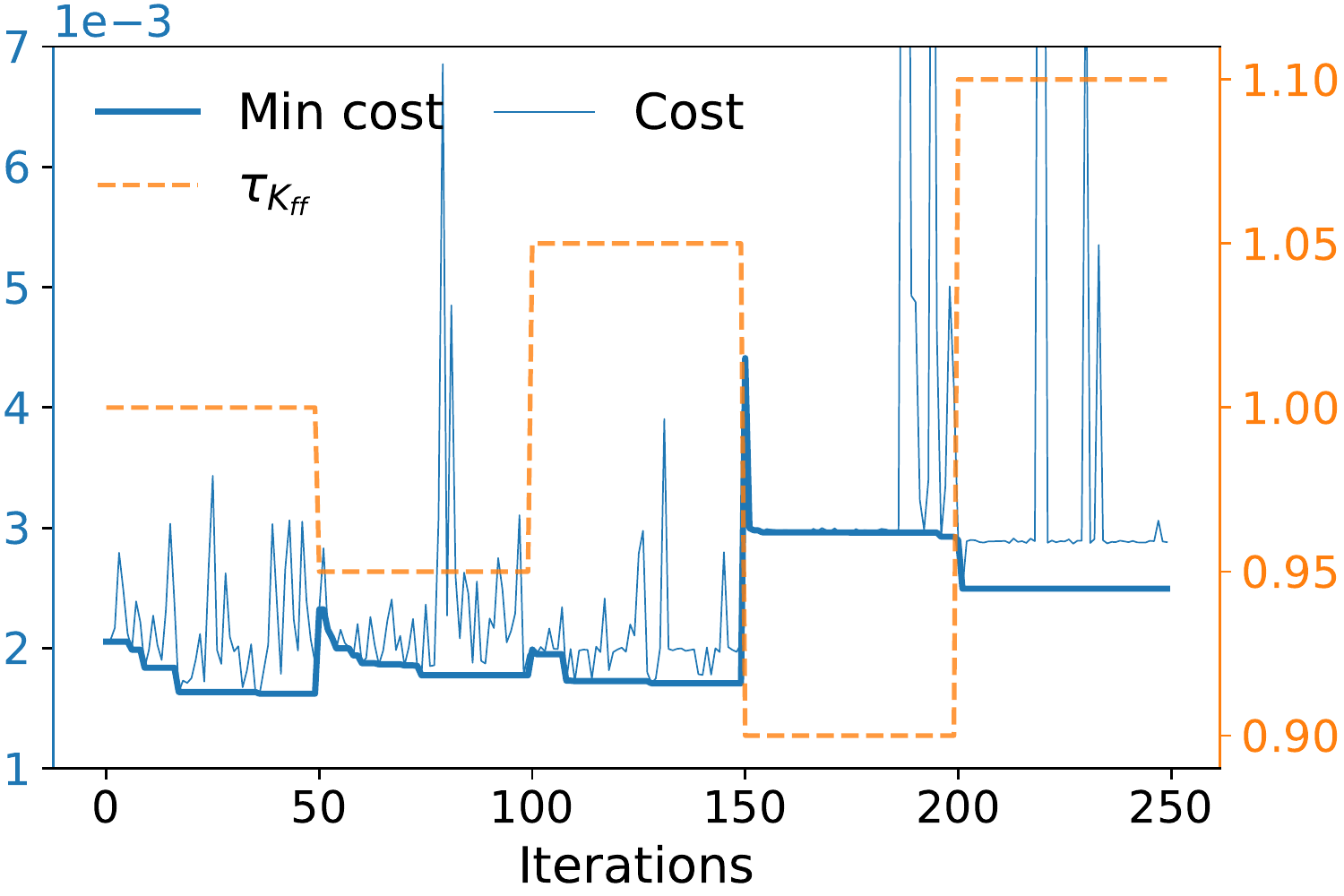}
\caption{Minimum and observed cost $f$.}
\label{fig:adaptiveC_exp_Kff_no_task_cost}
\end{subfigure}
\hfill
\begin{subfigure}[c]{0.25\textwidth}
\includegraphics[width=1\textwidth]{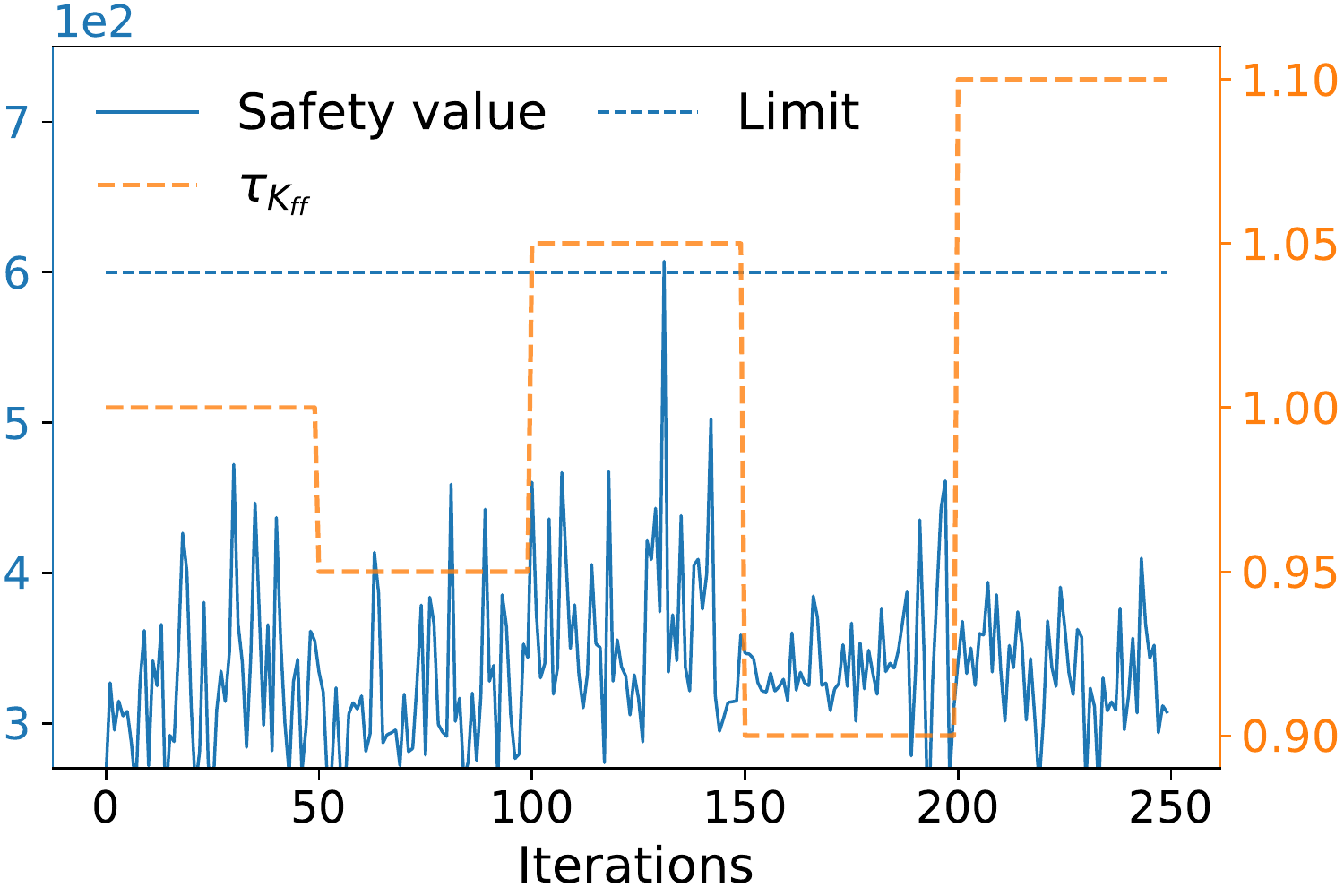}
\caption{Safety constraint $q_1$.}
\label{fig:adaptiveC_exp_Kff_no_task_q1}
\end{subfigure}
\hfill
\begin{subfigure}[c]{0.25\textwidth}
\includegraphics[width=1\textwidth]{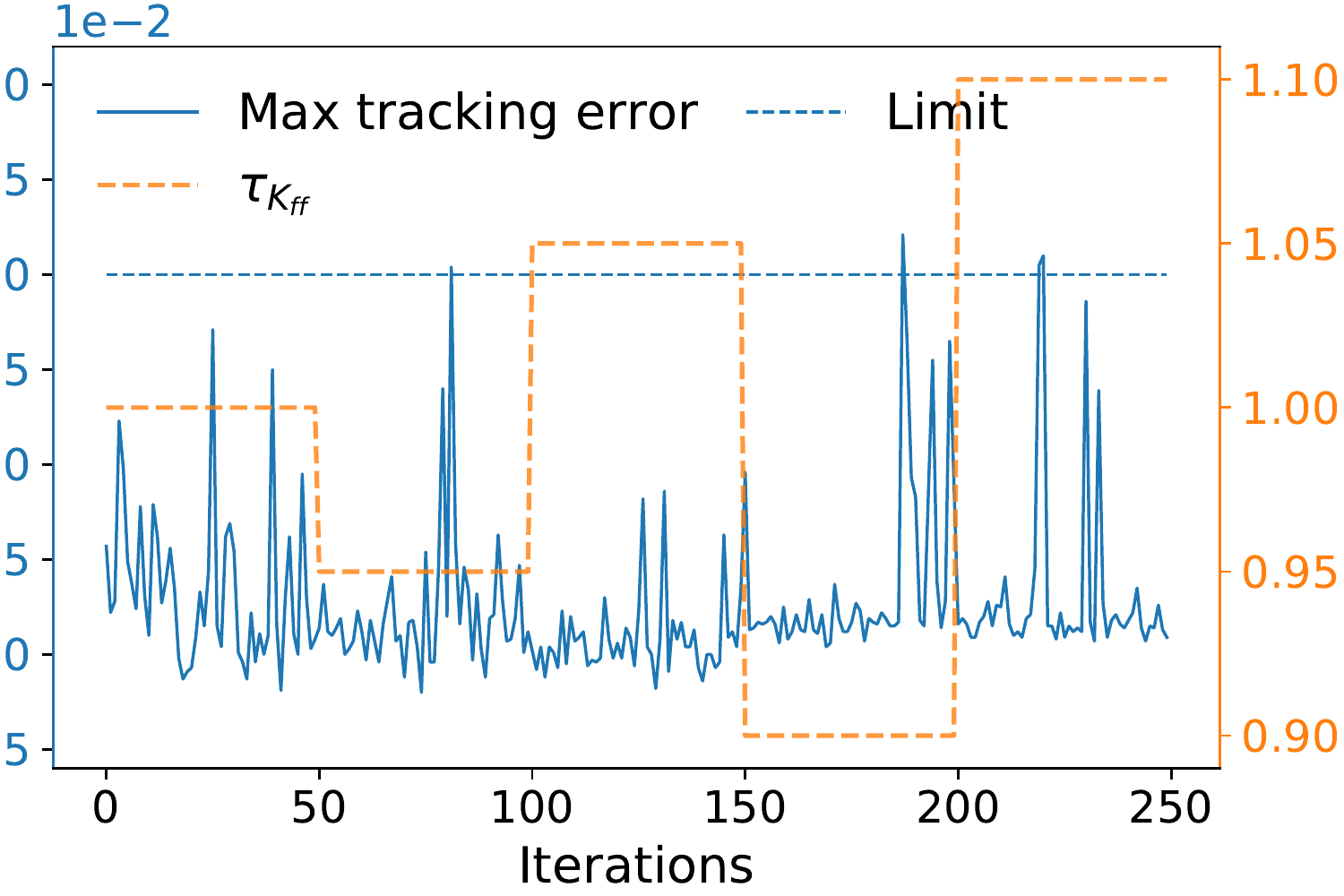}
\caption{Tracking error constraint $q_2$.}
\label{fig:adaptiveC_exp_Kff_no_task_q2}
\end{subfigure}
\hfill
\footnotesize\addtolength{\tabcolsep}{-5pt}
\raisebox{0.3cm}{
\begin{tabular}{ @{}l c c c c c c@{} }\toprule
$\task_{K_\text{ff}}$ & $1.0$ & $0.95$ & $1.05$ & $0.9$ & $1.1$\\ \midrule
$f(\xbest)$ & $1.62$ & $2.08$ & $2.01$ & $3$ & $2.89$\\
$\Kpbest$ & $25$ & $25$ & $25$ & $25$ & $25$\\
$\Kvbest$ & $.07$ & $.07$ & $.07$ & $.07$ & $.07$\\
$\Tibest$ & $3$ & $3$ & $3$ & $3$ & $3$\\
\bottomrule
\end{tabular}}}
\caption{\looseness=-1\small Cost (left), safety constraint (center), performance constraint (right) and summary (table) for the real-world adaptive control experiments with sudden change of the feed-forward gain $K_{\text{ff}}$, without considering the task parameter. The table summarizes the performance of the algorithm on the real system for 10 different experiments. Without keeping track of the task parameter, the cost for each different $K_{\text{ff}}$ after the first is higher, compared to task parameter tracking. While safety is mostly maintained, the performance constraint is violated.}
\label{fig:adaptiveC_exp_Kff_no_task}
\end{figure*}

\mypar{Experimental result} 
In comparison to the two numerical experiments of figure \cref{fig:adaptiveC_sim_no_task}, where we see that the algorithm fails in maintaining safety when we don't inform the GPs about a system relevant change, in this experiment we show that the performance can also deteriorate, without loosing stability. The internal change of $K_{\mathrm{ff}}$ has a marginal influence on the set of stable controller parameters. However, the table in \cref{fig:adaptiveC_exp_b} shows that it greatly influences the location of the optimal parameters within this set. Without tracking a task parameter representing the change in $K_{\mathrm{ff}}$, the algorithm optimizes the objective function for the initial value of $K_{\mathrm{ff}}$. Since there is no controller in later stages that outperforms the optimum of the first stage, the algorithm quickly reaches the termination criterion and applies the optimal controller found in the the first stage to the system. This results in sub-optimal costs as can be seen by comparing \cref{fig:adaptiveC_exp_b,fig:adaptiveC_sim_no_task} (left plots and tables). Furthermore, similar to the experiments of \cref{fig:adaptiveC_sim_no_task}, some parameter settings used by the algorithm after switching the initial $K_{\mathrm{ff}}$ value violate the safety or the performance constraint, since similar settings were accepted as safe in prior configurations/tasks.

\nomenclature{$\Kp$}{Position proportional gain}
\nomenclature{$\Kv$}{Velocity proportional gain}
\nomenclature{$\Ti$}{Velocity time constant}
\nomenclature{$\domain$}{Space of admissible controllers}
\nomenclature{$f$}{Objective (Average tracking error)}
\nomenclature{$q_1$}{Safety constraint (FFT of the torque)}
\nomenclature{$q_2$}{Performance constraint (Max tracing error)}
\nomenclature{$\mu$}{GP mean}
\nomenclature{$k$}{GP kernel}
\nomenclature{$\data$}{Data set}
\nomenclature{$\task$}{Task parameter}
\nomenclature{$l_t(x)$}{Lower bound of confidence interval}
\nomenclature{$u_t(x)$}{Upper bound of confidence interval}
\nomenclature{$S_0$}{Initial safe seed}
\nomenclature{$\g{x}{z}$}{Optimistic noisy expansion operator}
\nomenclature{$S^o$}{Optimistic safe set}
\nomenclature{$\suggestion$}{Recommendation of PSO}
\nomenclature{$x_w^*$}{Expander}
\nomenclature{$z_i$}{$i^{th}$ particle best position}
\nomenclature{$\overline{z}_i$}{Overall particles best position}
\nomenclature{$S$}{Safe set}
\nomenclature{$L$}{Safe set boundary}
\nomenclature{$L_q$}{Lipschitz constant}
\nomenclature{$W$}{Uncertain points on safe set boundary}
\nomenclature{$\xbest(\task)$}{Feasible controller corresponding to best observed cost for task $\task$}
\nomenclature{$\Kpbest$}{Position gain in $\xbest$}
\nomenclature{$\Kvbest$}{Velocity gain in $\xbest$}
\nomenclature{$\Tibest$}{Time constant in $\xbest$}
\nomenclature{$\xbestgrid(\task)$}{Best feasible controller computed with grid search for task $\task$}

\printnomenclature
 

\balance
\bibliography{bib}
\bibliographystyle{IEEEtran}


\end{document}